\documentclass[12pt,oneside]{book}
\usepackage{amsmath}
\usepackage{amssymb}

\def\be{\begin{equation}}
\def\ee{\end{equation}}
\def\bea{\begin{eqnarray}}
\def\eea{\end{eqnarray}}

\def\'#1{\if#1i{\accent 19\i}\else{\accent 19 #1}\fi}

\def\coes{\-\c c\~oes }

\def\e{\-men\-te }
\def\o{\-men\-to }

\def\cc{\c{c}}
\def\b{ }

\def\6{\partial} \def\a{\alpha} \def\b{\beta}
\def\g{\gamma} \def\d{\delta} \def\ve{\varepsilon}
\def\e{\epsilon} 
 \def\h{\eta} \def\th{\theta}
 \def\k{\kappa} \def\l{\lambda}
\def\m{\mu}   
 \def\s{\sigma} \def\t{\tau}
  
\def\o{\omega} \def\G{\Gamma} 
 \def\L{\Lambda} 
  \def\O{\Omega}
 \def\la{\large}

\def\non{\nonumber\\}
\def\ve{\vec}

\def\v{\upsilon}
  
\def\le{\left}
\def\ri{\right}

\def\bo{\boldmath}
\def\n{\nu}
\def\la{\langle}
\def\ra{\rangle}

\def\w{\widetilde}
\def\g{\gamma}

\newcommand{\beas}{\begin{eqnarray*}}
\newcommand{\eeas}{\end{eqnarray*}}
\newcommand{\bi}{\begin{itemize}}
\newcommand{\ei}{\end{itemize}}
\newcommand{\bc}{\begin{center}}
\newcommand{\ec}{\end{center}}
\newcommand{\bfl}{\begin{flushleft}}
\newcommand{\efl}{\end{flushleft}}
\newcommand{\bfr}{\begin{flushright}}
\newcommand{\efr}{\end{flushright}}
\newcommand{\f}{\frac}

\newcommand{\GG}{{\cal G}}
\newcommand{\HH}{{\cal H}}

\newcommand{\LL}{{\cal L}}

\newcommand{\OO}{{\cal O}}

\newcommand{\ZZ}{{\cal Z}}

\begin{document}

\title{
\vspace{4cm}
Contribui\cc o\~es ao Estudo dos Estados T\'ermicos da Corda Bos\^onica
no Formalismo de Din\^amica de Campos T\'ermicos}
\vspace{1cm}
\author{\Large Edison Luiz da Gra\cc a}
\maketitle

\newpage
\thispagestyle{empty}

\begin{center}
{\large Tese de Doutorado}
\end{center}
\vspace{4cm}
\begin{center}
{\Large\bf Contribui\cc o\~es ao Estudo dos Estados T\'ermicos da
Corda Bos\^onica no Formalismo de
Din\^amica de Campos T\'ermicos}
\end{center}
\vspace{1.5cm}
\begin{center}
{\Large Edison Luiz da Gra\cc a}
\end{center}
\vspace{8cm}
\begin{center}
{\large Centro Brasileiro de Pesquisas F\'{i}sicas\\
Janeiro de 2007}
\end{center}

\newpage
\thispagestyle{empty}

\begin{center}
{\bf Resumo}
\end{center}

Determinamos a entropia local e a energia livre para cordas t\'ermicas bos\^onicas abertas quantizadas no espa\cc o-tempo Minkowski, com as mais gerais condi\cc\~oes de contorno. Formulamos uma teoria a temperatura finita para as excita\cc\~oes t\'ermicas da corda bos\^onica fechada no espa\cc o-tempo anti-de Sitter, com abordagem da DCT. Escrevemos os estados e obtemos a entropia e a energia livre, com uma teoria perturbativa semicl\'assica quantizada at\'e primeira ordem, no referencial de centro de massa.

\vspace{1cm}

Palavras chave: Teoria de Cordas e Temperatura Finita.

\'Area de conhecimento: Teoria Qu\^antica de Campos.

\newpage
\thispagestyle{empty}

\begin{center}
{\bf Abstract}
\end{center}

We determine the local entropy of the free energy of the quantized open bosonic string in Minkowski spacetime with the most general boundary conditions. We formulate a finite temperature theory of the thermal closed string excitations in anti-de Sitter spacetime within the TFD approach. We write down the thermal states and obtain the entropy and the free energy in the first order expansion of the semiclassical quantization in the center of mass reference frame.

\newpage
\thispagestyle{empty}

\begin{center}
{\bf Agradecimentos}
\end{center}

Ao professor Ion Vasile Vancea, meu orientador, pelo coragem e perseveran\cc a que demonstrou nas horas dif\'iceis e que atrav\'es de uma conviv\^encia di\'{a}ria possibi- litou escrever esta tese. Ao professor Sebasti\~{a}o Alves
Dias, meu co-orientador, pela amizade e apoio. Ao professor Jos\'{e} Abdalla
Helay\"{e}l-Neto pelas excelentes aulas e semin\'{a}rios e pela sua amizade
sincera. Ao professor An\'{\i}bal Caride que  nos recebeu no CBPF. A Patricia
Vancea pelo acolhimento e carinho. A todos os conhecidos do CBPF, todos mesmo, que sempre
am\'{a}veis, foram sol\'{\i}citos \`as nossas necessidades. 

\newpage
\tableofcontents
\newpage

\chapter{Introdu\cc\~ao}

Os experimentos dispon\'iveis atualmente n\~ao atingem a escala de energia necess\'aria para testar a teoria de cordas,
entretanto o interesse na teoria de cordas se deve a possibilidade de que a teoria \'e uma forte candidata a unificar as for\cc as existentes na natureza. O estudo do espectro de cordas bos\^onicas, mostra que dependendo dos modos no qual a corda vibra, surgem part\'iculas que podem ser associadas a f\'otons e aos gr\'avitons, levando a pensar que a teoria \' e capaz de acomodar uma teoria qu\^antica da gravidade.

Recentemente, h\'a interesse na formula\cc\~ao das cordas e $D$-branas \`a temperatura finita por v\'arias raz\~oes.
A rela\c{c}\~{a}o entre as cordas e a teoria de campos \`{a} temperatura
finita representa por si mesma, um interessante problema que pode nos ajudar a
me\-lhor entender as propriedades f\'{\i}sicas das cordas e $D$-branas.  Algum
progresso nesta dire\c{c}\~{a}o pode ser feito no limite de baixas energias da
teoria de cordas, onde as $D$-branas s\~{a}o solu\c{c}\~{o}es solit\^{o}nicas
da (super)gravidade. Neste limite, a termodin\^{a}mica das cordas e $D$-branas
tem sido formulada utilizando as integrais de trajet\'{o}ria de teoria de
campos \`{a} temperatura finita \cite{mvm}-\cite{gub}.
De outro lado podemos querer entender as propriedades estat\'{\i}sticas de
alguns sistemas que podem ser descritos em termos de cordas, $D$-branas e
anti-$D$-branas, como, por exemplo, o extremo, quasi-extremo e buracos negros
de Schwarzschild.

No outro bem conhecido limite da teoria de cordas, o limite perturbativo, as
informa\c{c}\~{o}es geom\'{e}tricas relativas as D-branas s\~{a}o perdidas.
Neste caso, as D-branas s\~{a}o apropriadamente descritas por uma
superposi\c{c}\~{a}o de estados coerentes no espa\c{c}o de Fock do setor de
cordas fechadas \cite{gw}-\cite{dv2} que devem satisfazer um conjunto de
condi\c{c}\~{o}es de contorno de Dirichlet e Neummann a serem impostas nos
pontos terminais da corda aberta. A interpreta\c{c}\~{a}o intuitiva das
D-branas como estados co\-erentes de contorno \'{e} mantida a temperatura finita
se a abordagem DCT \'{e} aplicada. A raz\~{a}o para isso \'{e} que a
depend\^{e}ncia t\'{e}rmica \'{e} implementada atrav\'{e}s dos operadores
t\'{e}rmicos que preservam a forma das rela\c{c}\~{o}es a temperatura zero.
Trabalhando com a DCT em vez do formalismo de integrais de trajet\'{o}ria a
tempo real temos uma formula\c{c}\~{a}o conveniente do problema, \'{e}
conhecido que ambos formalismos s\~{a}o equi\-valentes no equil\'{\i}brio t\'{e}rmico.

A DCT\ foi usada para discutir um g\'{a}s ideal de cordas, construir uma
teoria de campos de cordas bos\^{o}nicas abertas a temperatura finita e provar sua renormalizabilidade \cite{yl1}-\cite{fn2}. Estes estudos 
foram motivados pela
necessidade de entender a cosmologia de cordas e conjuntos de cordas em geral.
Todavia quando aplicamos a DCT para as cordas e $D$-branas, devemos tomar
algumas precau\c{c}\~{o}es \cite{ng1}-\cite{ivv10}. As $D$-branas s\~{a}o definidas como
estados no espa\c{c}o de Fock das cordas em primeira quantiza\c{c}\~{a}o. A
teoria de campos conformes que baseia a constru\c{c}\~{a}o descreve o
v\'{a}cuo bos\^{o}nico da teoria de cordas. Considerando temperatura finita,
interpretamos a corda t\'{e}rmica como um modelo para as excita\c{c}\~{o}es
t\'{e}rmicas do v\'{a}cuo bos\^{o}nico na teoria de cordas. Al\'{e}m disso, a
DCT \'{e} aplicada na teoria conforme em duas dimens\~{o}es. Assim, as
$D$-branas podem ser interpretadas como estados t\'{e}rmicos coerentes de
contorno no espa\c{c}o de Fock das excita\c{c}\~{o}es t\'{e}rmicas \cite{ivv1}-\cite{ivv10}
.Uma outra nota \'{e} que a entropia das cordas e $D$-branas \'{e} definida
com o valor esperado do operador entropia \ no estado de v\'{a}cuo t\'{e}rmico
da corda.\ At\'{e} o presente n\~{a}o \'{e} co\-nhecida uma teoria na qual as
$D$-branas s\~{a}o escritas por estados do tipo v\'{a}cuo ou estados criados a
partir do v\'{a}cuo. Assim, para calcularmos a entropia dos estados de
$D$-branas temos que calcular o valor esperado do operador entropia da corda
bos\^{o}nica nos estados de contorno.

Usando a motiva\c{c}\~{a}o acima apresentada vamos construir os estados
t\'{e}rmicos da corda bos\^{o}nica e desenvolver um c\'{a}lculo para a
entropia tanto para cordas abertas no espa\c{c}o-tempo de Minkowski quanto
para cordas fechadas no espa\c{c}o-tempo anti-de Sitter (AdS). A corda
bos\^{o}nica no AdS \ representa o primeiro exemplo de quantiza\c{c}\~{a}o
exata da teoria de cordas no espa\c{c}o-tempo com curvatura. A diferen\c{c}a
entre a din\^{a}mica da corda nos espa\c{c}os de Minkowski e AdS \'{e} que em
geral AdS n\~{a}o \'{e} uma solu\c{c}\~{a}o das equa\c{c}\~{o}es de
fun\c{c}\~{o}es-$\beta$ para o modelo-$\sigma$ de corda. Portanto h\'{a} uma
grande classe de configura\c{c}\~{o}es de campos no AdS conformais e
n\~{a}o-conformais nas quais as propriedades f\'{\i}sicas das cordas
qu\^{a}nticas s\~{a}o dif\'iceis de estudar. Os fundos que s\~{a}o invariantes
conformais s\~{a}o necess\'{a}rios para definir a consist\^{e}ncia da teoria
qu\^{a}ntica de corda. Todavia muitos fundos interessantes do ponto de vista
f\'{\i}sico n\~{a}o satisfazem este requisito. Um m\'{e}todo para analisar a
din\^{a}mica de cordas bos\^{o}nicas no espa\c{c}o-tempo com m\'{e}trica
arbitr\'{a}ria foi proposto nos trabalhos \cite{vesan}-\cite{ns4}. Foi mostrado que
escolhendo as condi\c{c}\~{o}es apropriadas de contorno para a corda
bos\^{o}nica a invari\^{a}ncia de reparametriza\c{c}\~{a}o da teoria de folha
mundo pode ser escrita como uma transforma\c{c}\~{a}o de coordenadas entre
diferentes referenciais no espa\c{c}o-tempo. Tamb\'{e}m, um calibre de cone de
luz local pode ser escolhido em qualquer referencial. Neste calibre nos
podemos localmente separar os graus de liberdade da corda em longitudinais, ou
seja ao longo da trajetoria do centro de massa da corda, e transversais, e
mostrar que os graus de liberdade longitudinais s\~{a}o fun\c{c}\~{o}es
somente dos transversais. Esse permite um esquema de aproxima\c{c}\~{a}o para
a quantiza\c{c}\~{a}o can\^{o}nica em fundos invariantes conformais, esquema
esse chamado de \emph{quantiza\c{c}\~{a}o semicl\'{a}ssica}, no qual a
m\'{e}trica \'{e} tomada fixa enquanto a perturba\c{c}\~{a}o \'{e} feita em
torno da tra\-jetoria do centro de massa. Nos mesmos trabalhos o m\'{e}todo da
quantiza\c{c}\~{a}o semicl\'{a}ssica foi extendido at\'{e} primeira ord\'{e}m
para fundos AdS n\~{a}o-conformais D-dimensionais.

No cap\'{\i}tulo 2, desenvolvemos uma introdu\c{c}\~{a}o a cordas
bos\^{o}nicas no espa\c{c}o-tempo de Minkowski para cordas abertas e no
espa\c{c}o-tempo anti-de Sitter (AdS) \ o estudo para cordas fechadas.
Apresentam-se tamb\'{e}m, para ambas situa\c{c}\~{o}es as maneiras de proceder
para a devida quantiza\c{c}\~{a}o. No cap\'{\i}tulo 3 introduzimos o
formalismo da din\^{a}mica de campos t\'{e}rmicos (DCT). No quarto
cap\'{\i}tulo s\~{a}o a\-pre\-sentadas as contribui\c{c}\~{o}es mais relevantes
desta tese, analisamos a corda bos\^{o}nica aberta t\'{e}rmica no
espa\c{c}o-tempo de Minkowski, quantizada e calculamos a entropia local e a
energia livre para as mais variadas condi\c{c}\~{o}es de contorno
impostas. Ainda neste cap\'{\i}tulo consideramos a corda bos\^{o}nica fechada
quantizada no formalismo semicl\'{a}ssico no espa\c{c}o AdS, escrevemos os
estados f\'{\i}sicos e calculamos a entropia local e a energia livre. 
Por \'ultimo, discutimos a rela\c{c}\~{a}o entre a Hamiltoniana no espa\c{c}o de Hilbert 
total e o espa\c{c}o de Hilbert f\'{\i}sico. No \'{u}ltimo cap\'{\i}tulo s\~{a}o apresentadas as conclus\~{o}es e perspectivas futuras. A tese foi baseada nos trabalhos do autor \cite{eg1}-\cite{eg4}.

\newpage
\chapter{Corda Bos\^onica}

Apresentam-se neste cap\'{\i}tulo aspectos b\'asicos da teoria de cordas bos\^onicas tanto no espa\cc o-tempo de Minkowski quanto no espa\cc o AdS. Ser\~ao discutidas a a\cc \~{a}o cl\'assica de Polyakov e a sua quantiza\cc \~{a}o
can\^onica no calibre de cone de luz no espa\cc o de Minkowski \cite{green,polchinski}. A quantiza\cc \~{a}o semicl\'assica da corda no espa\cc o AdS \'e desenvolvida em \cite{vesan}-\cite{bhtz}.

\section{Corda Bos\^onica Cl\'assica no Espa\c{c}o - Tempo de Minkowski}

Uma superf\'{\i}cie bidimensional, denominada folha mundo \'e descrita pela corda ao propagar-se no espa\cc o-tempo. A folha mundo $M$ pode ser parametrizada pelas coordenadas $(\s^0 , \s^1 ) =( \t, \s )$, onde $\s^0 = \t$ \'e um par\^ametro tipo-tempo e o outro $\s^1 = \s \in [0, \pi ]$ \'e um par\^ametro tipo-espa\cc o. Uma fun\cc \~ao dessas coordenadas, para descrever a evolu\cc \~ao espa\cc o-temporal da corda na folha mundo $M$, \'e dada por 
$x^{a}(\s^0 , \s^1 )$, onde $a = 0, 1, \ldots , D-1$ e sendo $D$ a dimens\~ao do espa\cc o-tempo de Minkowski. 

A a\cc \~ao de Polyakov, que descreve a corda, \'e dada por
\be
S = -\frac{T_s}{2}\int d^{2}\s {\sqrt{-h}}h^{\a\b}\6_{\a}x^a \6_{\b} x^b \eta_{ab},  \label{acaopolyakov}
\ee
sendo $T_s =(2 \pi \a ')^{-1}$ e $\a'$ o par\^ametro de Regge. $h_{\a\b}$ e $\eta_{ab}$ s\~ao tensores m\'etricos de tipo-Minkowskiano, respctivamente, da folha mundo $M$ e do espa\cc o-tempo. $h_{\a\b}$ por ser sim\'etrico, possui tr\^es campos independentes, $h$ \'e o seu determinante e $d^2 \s = d \s^0 d \s^1$.  

A a\cc \~ao (\ref{acaopolyakov}) \'e invariante por transforma\coes gerais de coordenadas na folha\-mundo $\s^{\a} \rightarrow \s^{\a} + \xi^{\a}$. As reparametriza\cc\~oes locais sob as quais esta a\cc\~ao \'e invariante s\~ao
\bea
\d h^{\a\b}&=& \xi^{\g}\6_{\g}h^{\a\b}-\6_{\g}\xi^{\a}h^{\g\b}-\6_{\g}\xi^{\b}h^{\a\g}, \non
\d x^a&=&\xi^{\a}\6_{\a}x^a, \non
\d(\sqrt{-h})&=&\6_{\a}(\xi^{\a}\sqrt{-h}),
\label{simetriasPolyakov}
\eea

A a\cc\~ao (\ref{acaopolyakov}) apresenta invari\^ancia conforme ou de Weyl (por reescalamento conforme da m\'etrica $h_{\a\b}$):
$$\d x^a=0~~,~~\d h_{\a\b}=\L h_{\a\b},$$
onde $\L=\L(\s^0,\s^1)$ \'e uma fun\c{c}\~ao infinitesimal arbitr\'aria de $\s^{\a}$. Existe uma simetria global no espa\cc o-tempo de Minkowski, a a\cc\~ao \'e invariante de Poincar\'e para 
$$\d x^a= \o^{a}_{b}x^b + a^b~~,~~ \d h_{\a\b}=0,$$
onde $a^b$ \'e um vetor constante e $\o_{ab}=\h_{ac}\o^{c}_{b}$ \'e um tensor 
anti-sim\'etrico.

O tensor energia-momento \'e definido como
\be
T_{\a\b}=-\frac {2}{T_s} \frac {1}{\sqrt {-h}} \frac {\d S}{\d h^{\a\b}},
\label{defTensorEM}
\ee
e sua forma expl\'icita \'e
\be
T_{\a \b}= - \frac{1}{2}  h_{\a\b}h^{\g\d} \6_{\g}x^a \6_{\d}x_a  + \6_{\a} x^a \6_{\b} x_a.
\label{formulaTensorEM}
\ee
S\~ao v\'{\i}nculos da teoria cl\'{a}ssica: 
\be
\mbox{Tr} ( T_{\a\b}) =0~~ \mbox{e}~~ T_{\a\b}=0, 
\label{vinculos}
\ee
que devem tamb\'em ser satisfeitos pela teoria qu\^antica.

Adotaremos um calibre conveniente, que servir\'a para diminuiremos os graus de liberdade das vari\'aveis din\^amicas que aparecem explicitamente na a\cc\~ao. Escolhemos uma parametriza\cc\~ao da folha mundo, tal que $h_{\a\b}=e^\L \h_{\a\b}$, onde $\h_{\a\b}$ \'e a m\'etrica plana da folha mundo ($\h_{\a\b} = \mbox{diag}(-1,+1)$). Denomina-se $e^\L $ de fator conforme, $\L = \L (\t, \s )$. Substituindo este calibre conforme na a\cc\~ao
\be
S= -\frac{T_s}{2}\int d^2 \s \eta^{\a\b}\6_\a x^a \6_\b x_a ,
\label{acaoconforme}
\ee
resultando o tensor energia-momento 
\bea
T_{00}&=&T_{11}= \frac{1}{2} (\dot x^2 + x^{'2})=0, \non 
T_{10}&=&T_{01}=\dot x \cdot x^{'}=0. \label{vinc}
\label{TensorEMcalconf}
\eea
Explicitamente, neste calibre conforme temos o tensor energia-momento:
\bea
T_{00} &=& T_{11} = \frac{1}{2}\left( \dot{x}^2 + {x'}^2 \right) = 0\nonumber\\
T_{10} &=& T_{01} = \dot{x}\cdot x' =0.
\label{temexpl}
\eea
Tomando a varia\cc\~ao da a\cc\~ao de Polyakov (\ref{acaoconforme}) com rela\cc\~ao a $x^{a}$
\be
\d S = 0 = -T_s \int_{\6 M} d\t \left(n^\s \6_\s x^a \right)\d x_a 
- T_s \int_{M}d^2 \s \left(\6^\a \6_\a x^a \right)\d x_a ,
\label{varS}
\ee
onde $n^\s$ \'e um versor normal ao contorno $\6 M$. As duas parcelas da $\d S =0 $ devem se anular separadamente. Da segunda parcela
\be 
\6^\a \6_\a  x^{a}=0 ,
\label{eqKG}
\ee
s\~ao as equa\cc \~oes de movimento da corda; sendo equa\cc\~oes de Klein-Gordon em duas dimens\~oes, sem o termo de massa. Da primeira parcela resultam as condi\cc\~oes de contorno(c.c.). Para a corda fechada, as condi\cc\~oes de contorno peri\'odicas escolhidas s\~ao:
\be
x^{a}(\t,0)=x^{a}(\t,\pi). \label{cccf}
\ee
Para a corda aberta
\be
\left[ \6_{\s} x^{a} \d x_{a} \right]_{\s=0}^{\s=\pi}=0, 
\label{ccca}
\ee
h\'a v\'arias possibilidades de condi\cc\~oes de contorno. $\left. \6_\s x^a \right |_{0}^{\pi}=0$ s\~ao c.c. tipo Neumann (N) e $\left. \d x^a \right |_{\6M}=0$ s\~ao c.c. tipo Dirichlet (D). As c.c. tipo N e D podem ser aplicadas independentemente \`{a}s duas extremidades da corda aberta. As solu\cc\~oes das equa\cc\~oes de movimento expandidas em s\'erie de Fourier com as c.c. NN, DD, DN e ND, respectivamente, s\~ao  
\bea
x^a(\t,\s)&=& x^{a} +2\a^{'}p^a \t +i\sqrt{2\a^{'}}\sum _{n \neq 0}\frac{1}{n}\a^{a}_{n}e^{-in\t}\cos{n\s}, \label{solNN} \\
x^{a}(\t,\s)&=&\frac{c^{a}(\pi - \s)+d^{a}\s}{\pi} - \sqrt{2 \a^{'}} \sum_{n \neq 0} \le( \frac{\a^{a}_{n}}{n}e^{-in\t}\sin{n\s} \ri), \label{solDD}\\
x^{a}(\t,\s)& = &c^{a}- \sqrt{2 \a^{'}} \sum_{r \in {\it Z^{'}}} \le( \frac{\a^{a}_{n}}{n}e^{-in\t}\sin{n\s} \ri), \label{solND} \\
x^{a}(\t,\s) &= &d^{a}+ i\sqrt{2 \a^{'}} \sum_{r \in {\it Z^{'}}} \le( \frac{\a^{a}_{n}}{n}e^{-in\t}\cos{n\s} \ri),\label{solDN}
\eea 
onde $x^a$ e $p^a$ s\ ao coordenadas de posi\cc\~ao e momenta canonicamente conjugados do centro de massa da corda, $c^a$ e $d^a$ s\~ao vetores constantes que descrevem respectivamente, as posi\cc\~oes dos extremos finitos onde a corda \'e aberta e $Z' = Z + 1/2$. Somente a solu\cc\~ao (\ref{solNN}) \'e invariante de Poincar\'e, as demais solu\cc\~oes t\^em alguma extremidade fixa que associada a objeto f\'isico extenso d\'a origem a chamada $D$-brana \cite{polchinski}. Vamos de agora em diante, somente considerar para cordas abertas as solu\cc\~oes NN, n\~ao trataremos de branas.

Para a corda fechada, as solu\cc\~oes das equa\cc \~oes de movimento s\~ao invariantes de Poincar\'e e dadas por
\be
x^a(\t,\s)= x^{a} +2\a^{'}p^a \t+i\sqrt{2\a^{'}}\sum _{n \neq 0}\frac{1}{2n}\le( \a^{a}_{n}e^{-2in(\t - \s)}+{\b}^{a}_{n}e^{-2in(\t + \s)}\ri ). \label{solcordafechada} 
\ee
Estas solu\cc\~oes para as cordas bos\^onicas s\~ao uma superposi\cc\~ao linear de modos de oscila\cc\~ao movendo-se para a direita e para a esquerda da corda, com respectivamente, coeficiente de Fourier $\a^{a}_{n}$ e $\b^{a}_{n}$. O fato de $x^{a}(\t, \s)$ ser real imp\~oe:
\be
\a^{a}_{-n} = (\a^{a}_{n})^*~~,~~\b^{a}_{-n} = (\b^{a}_{n})^*,
\label{realidade}
\ee
para $n>0$. 

Fixando $\t$ a evolu\cc\~ao do sistema pode ser descrita com os parenteses de Poisson para as vari\'aveis din\^amicas do sistema cl\'assico
\bea
\{ x^{a}(\t,\s), x^{b}(\t,\s') \} &=& \{ \dot{x}^{a}(\t,\s), \dot{x}^{b}(\t,\s') \} =0
\label{Poisson1}\\
\{ p^{a}(\t,\s), x^{b}(\t,\s') \} &=& T_s \{ \dot{x}^{a}(\t,\s), \dot{x}^{b}(\t,\s') \} = \eta_{ab}\delta(\s-\s')
\label{Poisson2}
\eea
Substituindo a solu\cc\~ao para corda fechada na \'ultima rela\cc\~ao
\be
\{ \a^{a}_{m} , \a^{b}_{n} \} = \{ \b^{a}_{m} , \b^{b}_{n} \} = i m \d_{m+n,0}\eta^{ab}~~,~~\{ \a^{a}_{m} , \b^{b}_{n} \} =0,
\label{PoissonFourier}
\ee
com as vari\'aveis do centro de massa
\be
\{ p^{a} , x^{b} \} = \eta^{ab}.
\label{cmPoisson}
\ee

\'E conveniente utilizar as componentes do tensor energia-momento nas coordenadas de cone de luz na folha mundo $x^{\pm} = \t \pm \s$ e $\6_{\pm} = \frac{1}{2} ( \6_\t \pm \6_s )$. Assim
\bea
T_{++} &=& \frac{1}{2} \left( T_{00} + T_{01} \right) = \6_+ x^a \6_+ x_a , \nonumber\\ 
T_{--} &=& \frac{1}{2} \left( T_{00} - T_{01} \right) = \6_- x^a \6_- x_a.
\label{temcl}
\eea
As equa\cc\~oes de v\'inculos (\ref{vinculos}) tomam a sequinte forma
\be
T_{++} = T_{--} = 0,
\label{vinculosclMink}
\ee
valendo para as cordas abertas e cordas fechadas. As componentes de movimento para a direita e para a esquerda s\~ao, respectivamente
\bea
x^{a}_{R} (x^+ ,x^- ) &=& \frac{1}{2}x^a + \frac{1}{2}l^2p^ax^- \frac{i}{2}\sqrt{2\a'}
\sum_{n\neq 0} \frac{\a^{a}_{n}}{n}e^{-21nx^-},
\label{leftmode}\\
x^{a}_{L} (x^+ ,x^- ) &=& \frac{1}{2}x^a + \frac{1}{2}l^2p^ax^- \frac{i}{2}\sqrt{2\a'}
\sum_{n\neq 0} \frac{\a^{a}_{n}}{n}e^{-21nx^+}.
\label{rightmode}
\eea
Para a corda fechada, as componentes de Fourier de $T_{++}$ e $T_{--}$ definidas em $\t = 0$ s\~ao:
\bea
\bar{L}_m & = & \frac{T_s}{2}\int^{\pi}_{0}d\s e^{im\s}T_{++},
\label{barL}\\
L_m & = & \frac{T_s}{2}\int^{\pi}_{0}d\s e^{im\s}T_{--},
\label{L}
\eea 
Reescritos em modos de Fourier
\bea
\bar{L}_m & = & \frac{1}{2}\sum_{n=-\infty}^{\infty} \a^{a}_{m-n}\a_{an},
\label{barLFourier}\\
L_m & = & \frac{1}{2}\sum_{n=-\infty}^{\infty} \b^{a}_{m-n}\b_{an}.
\label{LFourier}
\eea 
$L_m$ e $\bar{L}_m$ s\~ao denominados de operadores de Virasoro. 

Para a corda aberta temos um conjunto de osciladores de modos $\a^{a}_{m}$ e definimos as componentes de Fourier do tensor energia-momento como
\be
L_m = T_s \int^{pi}_{0} d \sigma \left(e^{im\s }T_{++} + e^{-im\s}T_{--}\right)=
\frac{T_s}{2}\sum_{n=-\infty}^{\infty}\a^{a}_{m-n}\a_{an}.
\label{Laberta}
\ee

A Hamiltoniana para a corda aberta \'e $H=L_0$ e para a corda fechada $H = L_0 + \bar{L}_0$. Em termos de componentes de Fourier temos para a corda aberta:
\be
H= \frac{T_s}{2}\int d^s \left( \dot{x}^2 + {x'}^2 \right) = \sum _{n \neq 0} \a_{-n}^{a}\a_{a n} + \frac{1}{2}\sqrt{2\a^{'}}p^{a}p_{a}, \label{hca}
\ee
onde $\a_{0}^{a}=\sqrt{2\a^{'}}p^{a}$. Para a corda fechada a Hamiltoniana toma a seguinte forma 
\be
H=\sum _{n \neq 0} (\a_{-n}^{a}\a_{a n} + {\b}_{-n}^{a} {\b}_{a n})+\frac{1}{2}\sqrt{2\a^{'}}p^{a}p_{a}, \label{hcf}
\ee
onde $\a_{0}^{a}={\b}_{0}^{a}=\frac{1}{2}lp^{a}$.

A partir das defini\cc\~oes dos operadores de Virasoro e dos par\^enteses de Poisson dos modos de osciladores, escrevemos para corda aberta os par\^enteses de Poisson
\be
\{ L_m , L_n \} = i (m-n)L_{m+n},
\label{CAPoissonVirasoroClass}
\ee
para corda fechada
\bea
\{ L_m , L_n \} & = & i (m-n)L_{m+n}~~,~~ \{ \bar{L}_m , \bar{L}_n \} = i (m-n)\bar{L}_{m+n}, \nonumber\\
\{ L_m , \bar{L}_n \} & = & 0.
\label{CFPoissonVirasoroClass}
\eea

Utilizando a condi\cc\~ao de concha de massa $M=-p^a \cdot p_ a$ e o vinculo $ L_0 = 0$ para corda aberta, o quadrado da massa da corda $M^2$ \'e obtido em termos dos modos internos de oscila\cc\~ao
\be
M^2 = \frac{1}{\a'} \sum_{n=1}^{\infty} \a^{a}_{-n}\a_{an}.
\label{massaCA}
\ee
para corda fechada, os v\'inculos s\~ao $L_0 = \bar{L}_0 = 0$ e obtemos
\be
M^2 = \frac{2}{\a'} \sum_{n=1}^{\infty} \left( \a^{a}_{-n}\a_{an}+ \b^{a}_{-n}\b_{an} \right).
\label{massaCF}
\ee

\section{Quantiza\cc\~ao da Corda Bos\^onica no Cone de Luz}

Na escolha de calibre conforme nem toda liberdade de calibre foi removida,
ainda \'{e} poss\'{\i}vel reduzir o n\'{u}mero de componentes n\~{a}o triviais
de $x^{a}\left(  \tau,\sigma\right)  $ e que mant\'{e}m somente os graus de
liberdade fisicos relevantes \cite{ggrt}. Vamos definir as coordenadas de cone
de luz para uma corda em $D$ dimens\~{o}es
\be
x^{\pm}(\t,\s)=\frac{1}{\sqrt{2}}\left(x^{0}(\t,s) \pm x^{D-1}(\t,\s)\right). 
\label{conedeluz}
\ee 
A invari\^ancia residual de calibre permite fazer a escolha 
\be
x^{+}=x^{+}+l^{2} p^{+}\t ,
\label{labelxmais}
\ee
onde $x^+$ e $p^+$ s\~ao constantes. 

Combinando as reparametriza\cc\~oes e o reescalonamento local de Weyl podemos obter novo $\t$, que \'e soma de fun\cc\~oes arbitr\'arias de $(\t \pm \s)$. O novo $\t$ definido, pode ser identificado com qualquer solu\cc\~ao escolhida $u$ da equa\cc\~ao de onda
\be
\6^\a \6_\a u = 0.
\label{eqonda}
\ee
Como $x^+$ e $( ax^+ + b )$, para $a$ e $b$ constantes, satisfazem a equa\cc\~ao de onda (linearidade), a escolha (\ref{labelxmais}) \'e aceit\'avel para novo $\t$.

Considerando a corda aberta, suas componentes $x^{\pm}(\t,\s)$ satisfazem as mesmas solu\cc\~oes
(\ref{solNN})-(\ref{solDN}) com $a \rightarrow i = 1, 2, \ldots , D-1$. As rela\cc\~oes de quantiza\cc\~ao no cone de luz a serem satisfeitas pelos campos de corda s\~ao:
\be
[x^{i}(\t,\s ), P^{j}_{\t}(\t, \s^{'})]= i \d^{ij}\d(\s-\s^{'}),
\label{rel1}
\ee
\be
[x^{-}, p^{+}]=-i,
\label{rel2}
\ee
\be
[x^{i}(\t,\s ), x^{j}(\t, \s^{'})]=[ P^{i}_{\t}(\t, \s), P^{j}_{\s}(\t, \s^{'})]=0,
\label{rel3}
\ee
\be
[x^{-}, x^{i}]=[x^{-}, P^{j}_{\t}]=[p^{+}, x^{i}]=[p^{+}, P^{i}_{\t}]=0.
\label{rel4}
\ee
Das rela\cc\~oes (\ref{rel1})-(\ref{rel4}) segue que os operadores no espa\cc o de Fock para osciladores quantizados devem satisfazer 
\be
[\a^{i}_{n },\a^{j}_{m }]= n\d^{ij}\d_{n+m}.   \label{rcc}
\ee
\be 
\a^{i}_{-n}= (\a^{i}_{n})^\dagger,~~ n>0,
\ee 
O operador de massa qu\^antico pode ser escrito
\be 
M^{2} = \frac{1}{\a'}(N - 1),  \label{omca}
\ee
onde 
\be
N=\sum^{\infty}_{m=1}\a^{i}_{-m}\a^{i}_{m}.
\label{oneca} 
\ee
O Hamiltoniano no cone de luz para a corda aberta \'e 
\be
H_{ca}=\frac{1}{2}\sum^{\infty}_{m= -\infty}:\a^{i}_{m}\a^{i}_{-m}: - 1,
\label{HamCL}
\ee
e os operadores de Virasoro quantizados
\be
L_m = \frac {1}{2} \sum ^{\infty}_{-\infty} : \a_{m-n} \cdot \a_{n}:.
\label{Virasoroca}
\ee

A \'algebra de Virasoro para o caso qu\^antico n\~ao apresenta anomalias para $D =26$. Neste caso a \'algebra satisfaz
\be
\left [L_m,L_n \right ]= (m-n)L_{m+n}. \label{aqv}
\ee   
Em geral, a presen\c{c}a de anomalias na \'{a}lgebra qu\^{a}ntica de Virasoro
n\~{a}o permite que o v\'{\i}nculo cl\'{a}ssico $L_{m}=0$, $\forall m$ possa
ser implementado em estados qu\~{a}nticos. Por causa das rela\c{c}\~{o}es de
comuta\c{c}\~{a}o $[\alpha_{m}^{a},\alpha_{n}^{b}]=m\delta_{m+n,0}\eta^{ab}$
que definem o espa\c{c}o de Fock com os osciladores, conterem a m\'{e}trica de
Lorentz $\eta^{ab}$, existem no espa\c{c}o de Fock estados com norma negativa
(estados fantasma ou tamb\'{e}m chamados estados n\~{a}o-f\'{\i}sicos). No
formalismo de cone de luz s\~{a}o resolvidos os v\'{\i}nculos cl\'{a}ssicos e
o espa\c{c}o de Fock cont\'{e}m apenas os estados f\'{\i}sicos. 
O estado de v\'acuo com momento $p$ \'e definido como  
\bea
\a^{i}_{n}|0;p\rangle _{\a} &=& 0,~~ n>0,
\label{vacuoca1} \\
\hat{p}^{i}|0;p\rangle_{\a} & =& p^{i}|0;p\rangle_{\a}.
\label{vacuoca2} 
\eea
Como componentes de Fourier de $T_{ab}=0$, a Hamiltoniana $H = L_0$ e $L_m$ s\~ao as demais componentes para $m>0$, na corda aberta. Para os demais estados f\'isicos 
\be
L_{m}|\Psi_{phys} \ra=0 ~~,~~ m>0.
\label{vincca}
\ee
O operador $L_m$ tem a propriedade de hermiticidade 
\be 
L_{-m}=L_{m}^{\dagger}~~. 
\label{herma}
\ee

Considerando a corda fechada, temos duas \'algebras para os osciladores qu\^antizados
\be 
[\a^{i}_{n},\a^{j}_{m}] = [\b^{i}_{n},\b^{j}_{m}]=n\d^{ij}\d_{n+m}~~,~~
[\a^{i}_{n},\b^{j}_{m}] = 0. \label{osccf}
\ee
A Hamiltoniana da corda fechada \'e
\be
H_{cf} = \sum _{n=-\infty}^{\infty}
\le( :{\a}^{i}_{n}{\a}^{i}_{-n}:+:{\b}^{i}_{n}{\b}^{i}_{-n}: -2 \ri ), 
\label{Hamcf}
\ee
e o operador de massa \'e
\be
M^{2}=\frac{1}{\a'}(N + \bar{N}-2),
\label{omcf}
\ee
onde
\be
N=\sum _{n=1}^{\infty}{\a}^{i}_{-n}{\a}^{i}_{n}~~,~~ \bar{N}=\sum _{n=1}^{\infty}{\b}^{i}_{-n}{\b}^{i}_{n}.
\ee
Os operadores de Virasoro para a corda fechada s\~ao
\bea
L_{m}& = & \frac {1}{2} \sum^{\infty}_{n=-\infty} : \a_{m-n}^{a} \a_{a n}:~,~~ m \neq 0,
\label{Virasorocf1}\\
{\overline {L}}_{m} & = & \frac {1}{2} 
\sum^{\infty}_{n=-\infty} : {\b}_{m-n}^{a} {\b}_{a n}:~,~~ m \neq 0.
\label{Virasorocf2}
\eea
Os estados f\'{\i}sicos devem satisfazer a condi\c{c}\~ao de Virasoro
\bea
(L_{m}-\d_{m})|\Psi_{phys} \ra &=& 0 ~~,~~ m\geq 0 ,
\label{vincvir1}\\
(\overline{L}_{m}-\d_{m})|\Psi_{phys} \ra &=& 0~~,~~ m \geq 0 .
\label{vincvir2}
\eea
Os operadores de Virasoro para a corda fechada tem a propriedade de hermiticidade
\be 
L_{-m}=L_{m}^{\dagger}~~,~~ \overline{L}_{-m}=\overline{L}_{m}^{\dagger}. 
\label{hermf}
\ee
O v\'inculo cl\'assico $L_{0}=\overline{L}_{0}=0$ \'e implementado em (\ref{vincvir1}) e (\ref{vincvir2}) para $m=0$ 
\be
(L_{0}-1)|\Psi_{phys} \ra=(\overline{L}_{0}-1)|\Psi_{phys} \ra=0. \label{vcf}
\ee
e finalmente o v\'acuo na corda fechada satisfaz
\be
\a^{i}_{n}|0\rangle_{\a}|0\rangle_{\b} ={\b}^{i}_{n}|0\rangle_{\a}|0\rangle_{\b} =0,~~n > 0.    \label{vaccf} 
\ee

A obten\cc\~ao dos demais estados f\'isicos faz-se com a atua\cc\~ao sucessiva dos o\-pe\-ra\-dores de cria\cc\~ao de osciladores da corda sobre o estado de v\'acuo.

\section{Espa\c{c}o-Tempo AdS com $D=2+1$}

O espa\cc o AdS com $D=2+1$ pode ser embebido no espa\cc o com $D=2+2$ e com a m\'etrica \cite{bhtz,bhtz1}
\be
ds^2=-du^2-dv^2+dx^2+dy^2,
\label{3.1}
\ee
atrav\'es da equa\cc\~ao
\be
-v^2-u^2+x^2+y^2=-l^2.
\label{3.2}
\ee
Podemos definir um sistema de coordenadas para a variedade inteira
\be
u=l\cosh\mu \sin\lambda, \;\;\; v=l\cosh\mu \cos\lambda ,
\label{3.3}
\ee
onde $l\sinh\mu=\sqrt{x^2+y^2}$ e $0\leq \mu <\infty$,
$0\leq\lambda <2\pi$.  Usando as rela\cc\~oes (\ref{3.3}) e (\ref{3.1}) podemos escrever a m\'etrica da seguite forma
\be
ds^2= l^2\left(- \cosh^2\mu \ d\lambda^2 + \frac{dx^2+dy^2 }{l^2 +x^2
+y^2}\right).
\label{3.4}
\ee
Esta rela\cc\~ao pode ser simplificada em coordenadas polares no plano $(x,y)$
\be
x=l\sinh\mu \cos \theta, \;\;\; y=l\sinh \mu \cos \theta ,
\label{3.5}
\ee 
para obter a seguinte m\'etrica no espa\cc o AdS
\be
ds^2= l^2\left[ -\cosh^2 \mu d\lambda^2 + d \mu^2 + \sinh^2 \mu
d\theta^2 \right].
\label{3.6}
\ee
O par\'ametro $\lambda$ sendo um \^angulo existem curvas fechadas no espa\cc o AdS,
por exemplo $\mu=\mu_{0}, \theta =\theta_{0}$. Por esta raz\~ao n\~ao vamos identificar 
$\lambda$ com $\lambda+2\pi$. Usando as nota\cc\~oes $\lambda = t/l$ e $r=l\sinh\mu$ escrevemos (\ref{3.6}) na seguinte forma
\be
ds^2= ((r/l)^2+1)dt^2 + ((r/l)^2+1)^{-1}dr^2 +r^2d\theta^2.
\label{3.6.5}
\ee

A m\'etrica do espa\cc o AdS \'e invariante por constru\cc\~ao \`a a\cc\~ao do grupo
$SO(2,2)$. Os vetores de Killing s\~ao 
\be
J_{a b}=x_b \frac{\6}{\6 x^a} - x_a
\frac{\6}{\6 x^b},
\label{3.7}
\ee
onde $x^a=(v,u,x,y)$. A forma detalhada dos vetores (\ref{3.7}) \'e 
\be
\begin{array}{rclcrcl}
J_{01} &=& v\6_u - u \6_v \;\;&\; J_{02} &=& x\6_v
+ v \6_x, \\
J_{03} &=& y\6_v + v \6_y \;\;&\; J_{12} &=& x\6_u
+ u \6_x, \\
J_{13} &=& y\6_u + u \6_y \;\;&\; J_{23} &=& y\6_x
- x \6_y.
\end{array}
\label{3.8}
\ee
O vetor $J_{01}$ gera "transla\cc\~oes temporais" enquanto o vetor $J_{23}$ gera rota\cc\~oes no plano $(x,y)$. A forma mais geral do vetor de Killing \'e
\be
\m \omega^{ab} J_{ab},\;\; \;\;\; \omega^{ab}=-\omega^{ba},
\label{3.9}
\ee
sendo este determinado pelo tensor antisim\'etrico do espa\cc o ${\bf R}^4$.

Podemos definir as coordenadas de Poincar\'e atrav\'es das seguintes rela\cc\~oes
\be
z=\frac{l}{u+x}, \;\;\; \beta=\frac{y}{u+x}, \;\;\;
\gamma=\frac{-v}{u+x}.
\label{3.10}
\ee
Estas coordenadas cobrem apenas uma parte do espa\cc o AdS, ou seja uma infinidade de regi\~oes onde $u+x$ tem um sinal bem definido. Consequentemente, as coordenadas de Poincar\'e n\~ao s\~ao apropriadas para estudar as propriedades globais do AdS. Em fun\cc\~ao das $(z,\beta,\gamma)$ 
o elemento de linha do espa\cc o AdS tem a seguinte express\~ao
\be
ds^2 = l^2 \left[ \frac{ dz^2 + d\beta^2 - d\gamma^2 }{z^2}
\right].
\label{3.11}
\ee
Para $u+x>0$ temos $z>0$ enquanto para $u+x<0$ temos $z<0$. De forma semelhante podemos definir as coordenadas de Poincar\'e para cada regi\~ao onde $u-x$ tem um sinal definido.

Os vetores de Killing definem subgrupos uniparam\'etricos de isometrias do espa\cc o AdS 
\be
P \rightarrow e^{t\xi} P.
\label{3.12}
\ee
Os valores do $t$ n\'umero inteiro m\'ultiplo de $2\pi$
\be
P\rightarrow e^{t\xi}P, \;\;\;\;\;  t=0,\pm 2\pi,\pm 4\pi,....,
\label{3.13}
\ee
definem um {\em subgrupo de identifica\cc\~oes}. O espa\cc o quociente obtido atrav\'es da identifica\cc\~ao dos pontos de uma \'orbita dada do subgrupo de identifica\cc\~oes tem a m\'etrica de curvatura negativa induzida pela m\'etrica do AdS. Consequentemente, o espa\cc o
quociente \'e uma solu\cc\~ao das equa\cc\~oes de Einstein. A condi\cc\~ao necess\'aria para a aus\^encia de curvas fechadas do tipo tempo \'e  
\be
\xi \cdot \xi >0.
\label{3.14}
\ee
A rela\cc\~ao (\ref{3.14}) torna o vetor de Killing $\xi$ um vetor do tipo espa\cc o. Para buracos negros, esta condi\cc\~ao \'e tamb\'em suficiente. Existem vetores de Killing que satisfazem a rela\cc\~ao (\ref{3.14}) no espa\cc o inteiro. Contudo, alguns dos vetores de Killing que determinam a estrutura dos buracos negros s\~ao do tipo nulo ou temporal em certas regi\~oes do espa\cc o AdS. Estas regi\~oes devem ser recortadas do espa\cc o para fazer as identifica\cc\~oes poss\'{\i}veis. O espa\cc o resultante, chamado de ads,  \'e invariante \`{a}s transforma\cc\~oes (\ref{3.12}) porque a norma dos vetores de Killing \'e constante ao longo de suas \'orbitas. O espa\cc o ads \'e geodesicamente incompleto porque ele tem geodesicas que ligam $\xi \cdot \xi >0$ ao  $\xi \cdot
\xi <0$. As fronteiras da regi\~ao $\xi \cdot
\xi >0$, ou seja a superf\'{\i}cie $\xi \cdot \xi =0$, aparece como sendo uma singularidade na estrutura do espa\cc o-tempo e produz curvas do tipo tempo fechadas. Devido a esse fato, 
a regi\~ao com $\xi \cdot \xi =0$ pode ser vista como uma singularidade do espa\cc o quociente. Consequentemente, as \'unicas geodesicas incompletas s\~ao as que atingem a singularidade, como no caso dos buracos negros em $D=3+1$. A superf\'{\i}cie $\xi \cdot \xi =0$ \'e singular somente na estrutura causal \cite{bhtz}. 

A a\cc\~ao da gravita\cc\~ao no formalismo lagrangiano em unidades $G = \frac{1}{8}$ \'e
\be
I= \frac{1}{2\pi} \int \sqrt{-g} \left[ R + 2l^{-2}
\right]d^2 xdt  + B',
\label{1}
\ee
onde $B'$ \'e um termo de superf\'{\i}cie e o raio $l$ \'e relacionado \`a
constante cosmologica
$-\Lambda = l^{-2} $. A varia\cc\~ao da a\cc\~ao em rela\cc\~ao a m\'etrica
$g_{a b}(x,t)$ conduz \`as equa\cc\~oes de Einstein
\be
R_{a b} - \frac{1}{2} g_{a b} (R + 2 l^{-2}) = 0.
\label{2}
\ee
Em $D=1+2$ o tensor de Riemann \'e completamente determinado pelas rela\cc\~oes (\ref{2})
\be
R_{a b c d} = -l^{-2} (g_{a c} g_{b d}
- g_{b c} g_{a d}).
\label{3}
\ee
A rela\cc\~ao acima descreve um espa\cc o sim\'etrico de curvatura constante e negativa.
Para obter a solu\cc\~ao de buraco negro usamos o seguinte {\em Ansatze} \cite{bhtz}
\begin{equation}
ds^2=-a(r)dt^2+\frac{dr^2}{a(r)}+r^2 d\phi^2,
\end{equation}
onde $a(r)$ \'e uma fun\cc\~ao arbitr\'aria de $r$. O tensor de Einstein 
$G_{a b}=R_{a b}-\frac{R}{2}g_{a b}$ \'e
\begin{equation}
G_{rr}=\frac{a_{,r}}{2ar},\;\;\;G_{tt}=-\frac{aa_{,r}}
{2r},\;\;\;G_{\phi\phi}=
\frac{r^2}{2}a_{,rr},
\end{equation}
as outras componentes sendo nulas.
A \'unica solu\cc\~ao de v\'acuo \'e $a=\mbox{constante}$ e corresponde ao espa\cc o plano em $D=2+1$. Levando em considera\cc\~ao a constante cosmologica da a\cc\~ao (\ref{1})
\begin{equation}
T^\mu\;_\nu=\mbox{diag}(1,1,1)\Lambda,
\end{equation}
a solu\cc\~ao \'e n\~ao trivial
\begin{equation}
a(r)=c-\Lambda r^2,
\end{equation}
onde  $c$ \'e uma constante arbitr\'aria. Se $c=1$ obtemos duas solu\cc\~oes:
$\Lambda > 0$ representa o espa\cc o de Sitter, e $\Lambda < 0$
representa o espa\cc o anti de Sitter. Se $c < 0$ obtemos a solu\cc\~ao 
\cite{bhtz}
\begin{equation}
ds^2=(1-\frac{r^2}{l^2})dt^2+(\frac{r^2}{l^2}-1)^{-1}dr^2+r^2 d\phi^2.
\end{equation}
Esta solu\cc\~ao particular com $\phi \cong \phi+2\pi$ descreve um buraco negro de massa  $M=1$
e momento \^angular $J=0$. O horizonte est\'a localizado no $r=l$ e asimptoticamente a solu\cc\~ao tende ao espa\cc o AdS com $\Lambda=-1/l^2$. Uma familia de solu\cc\~oes biparam\'etricas em $(M,J)$ de buracos negros pode ser obtida identificando uma combina\cc\~ao linear de $t$ e $\phi$ o que leva \`a solu\cc\~ao \cite{bhtz}
\begin{equation}
ds^2=(M-\frac{r^2}{l^2})dt^2+(\frac{r^2}{l^2}-M+\frac{J^2}{4r^2})^{-1}dr^2
-Jdtd\phi+r^2 d\phi^2,\label{sol-int}
\end{equation}
com dois horizontes para $Ml^2>J^2$
\begin{equation}
r_\pm=\sqrt{\frac{Ml^2}{2}\pm\frac{l}{2}\sqrt{M^2 l^2-J^2}}
\end{equation}
e o limite est\'atico
\begin{equation}
r_{\mbox{erg}}=\sqrt{M}l,
\end{equation}
que define uma ergoesfera como para os buracos negros de Kerr.

Concluiremos esta se\cc\~ao fazendo algumas observa\cc\~oes sobre a relev\^ancia da solu\cc\~ao (\ref{sol-int}) para a teoria de cordas \cite{hw}. A a\cc\~ao da corda em primeira ordem em 
$\alpha'$ \'e
\begin{equation}
S=\int d^3 x\sqrt{-g}e^{-2\Phi}[\frac{4}{k}+R+4(\nabla\Phi)^2-\frac{1}{12}
H_{abc}H^{abc}],
\end{equation}
onde $\Phi$ \'e o campo dilat\^onico e $H_{abc}=\partial_{[a}
B_{bc]}$, sendo $B$ o campo de Kalb-Ramond. A m\'etrica (\ref{sol-int}) \'e uma solu\cc\~ao das equa\cc\~oes de movimento e da condi\cc\~ao \cite{hw}:
\begin{equation}
B_{\phi t}=\frac{r^2}{l^2},\;\;\;\Phi=0,\;\;\;k=l^2.
\end{equation}
Como foi mostrado em \cite{hw}, a rela\cc\~ao com o modelo-$\sigma$ pode ser feita atrav\'es da dualiza\cc\~ao em coordenada c\'{\i}clica $\phi$. A solu\cc\~ao dual \'e 
\cite{hw}
\begin{eqnarray}
&d\tilde{s}^2=(M-\frac{J^2}{4r^2})dt^2+(\frac{r^2}{l^2}-M+\frac{J^2}{4r^2})^{-1}
dr^2+\frac{2}{l}dtd\phi+\frac{d\phi^2}{r^2},&\nonumber
\end{eqnarray}
\begin{eqnarray}
&\tilde{B}_{\phi t}=\frac{-J}{2r^2},\;\;\;\tilde{\Phi}=-\log r.&
\end{eqnarray}
Ap\'os a diagonaliza\cc\~ao da m\'etrica obtemos
\begin{eqnarray}
&d\tilde{s}^2=-(1-\frac{{\cal M}}{\tilde{r}})d\tilde{t}^2+(1-\frac{{\cal Q}^2}
{{{\cal M}}\tilde{r}})d\tilde{x}^2+(1-\frac{{\cal M}}{\tilde{r}})^{-1}
(1-\frac{{\cal Q}^2}{{{\cal M}}\tilde{r}})^{-1}\frac{l^2 d\tilde{r}^2}
{4\tilde{r}^2},&\nonumber
\end{eqnarray}
\begin{eqnarray}
&\tilde{B}_{\tilde{x}\tilde{t}}=\frac{{\cal Q}}{r},\;\;\;
\tilde{\Phi}=-\frac{1}{2}
\log\tilde{r}l,&
\label{sol-final}
\end {eqnarray}
onde
\begin{eqnarray}
&t=\frac{l(\tilde{x}-\tilde{t})}{\sqrt{r^2_+-r^2_-}},\;\;\;\phi=\frac{r^2_+
\tilde{t}-
r^2_-\tilde{x}}{\sqrt{r^2_+-r^2_-}},&\nonumber
\end{eqnarray}
\begin{eqnarray}
&{\cal M}=\frac{r^2_+}{l},\;\;\;{\cal Q}=\frac{J}{2},\;\;\;r^2=\tilde{r}l.&
\end{eqnarray}
A m\'etrica (\ref{sol-final}) representa a solu\cc\~ao de corda negra em $D=2+1$ \cite{hw} obtida atrav\'es da fixa\cc\~ao de calibre do modelo-$\sigma$ com o grupo
$SL(2,R)\times R$.

\section{Corda Bos\^onica Cl\'assica no Espa\c{c}o-Tempo AdS}

A a\cc \~ao da corda bos\^onica no espa\cc o-tempo AdS \'e dada pelo funcional:
\be
S = \frac{1}{2 \pi \a '} \int d^2 \s \sqrt{h} h^{\a \b} g_{ab}(x) \6_{\a}x^{a} \6_{\b}x^{b}, 
\label{action}
\ee
escrevendo o tensor energia-momento
\be
T_{\a \b} \equiv \frac{2}{\sqrt{h}}\frac{\delta S}{\delta h^{\a \b}} = g_{ab}(x)\left( \6_{\a}x^{a} \6_{\b}x^{b} - \frac{1}{2} h_{\a \b} \6_{\g}x^{a} \6^{\g}x^{b} \right),
\label{energymomentum}
\ee
onde usamos
\be
\frac{\6 \sqrt{h}}{\6 h^{\a\b}} = \frac{1}{2}\sqrt{h}h_{\a\b}.
\ee
Pelas equa\cc\~oes de movimento de $h^{\a\b}$ o tensor energia-momento $T_{\a\b}=0$.
Com a fixa\cc\~ao do calibre conforme
\be
h_{\a \b}(\s^0 \s^1 ) = e^{\L(\s^0 , \s^1 )}\eta_{\a \b}, 
\label{conformalgauge}
\ee 
variando a a\cc\~ao com rela\cc\~ao a $x^{a}(\t,\s)$ e impondo $\d S = 0$, obtemos as equa\cc\~oes de movimento e os v\'inculos \cite{vesan}
\bea
\ddot{x}^{a} - x''^{a} + \G^{a}_{bc}(x)\left( \dot{x}^b\dot{x}^c - x'^{b}x'^{c} \right)&=&0,
\label{eqmotion}\\
g_{ab}(x)\dot{x}^a x'^{b} = g_{ab}(x)\left( \dot{x}^a\dot{x}^b + x'^{a}x'^{b} \right)&=&0,
\label{constraints}
\eea 
O m\'etodo de quantiza\cc\~ao semicl\'assica foi desenvolvido para estudar as excita\cc\~oes qu\^anticas em configura\cc\~oes cl\'assicas exatas (background). Devido a n\~ao-linearidade de (\ref{eqmotion})-(\ref{constraints}), vamos expandir as coordenadas $x^{a}(\t , \s )$ em torno de uma solu\cc\~ao exata $\eta^{a}_{0}(\t )$
\be
x^{a}(\t , \s ) = \sum^{\infty}_{n = 0}\e^n \eta^{a}_{n}(t, \s),
\label{fieldexpansion}
\ee
com a condi\cc \~ao inicial $\eta^{a}_{0}(\t , \s ) = \eta^{a}_{0}(\t )$ nas equa\cc\~oes de movimento e v\'inculos 
\bea
\ddot{\eta}^{a}_{0} + \G^{a}_{bc}(\eta_{0})\dot{\eta}^{b}_{0}\dot{\eta}^{c}_{0} &=& 0,
\label{eqmotioncm}\\
g_{ab}(\eta_0)\dot{\eta}^a_{0} \dot{\eta}^{b}_0 &=& -m^2 {\a '}^2.
\label{constraintcm}
\eea 

No espa\cc o AdS $D$ dimensional h\'a $D-1$ polariza\cc\~oes de perturba\cc\~oes da corda em torno da solu\cc\~ao $\eta^{a}_{0}(\t )$. Consequentemente, $D-1$ vetores normais transversos
$n^{a}_{\m}$, $\m = 1, 2, \ldots , D$ podem ser introduzidos
\bea
g_{ab}(\eta_0 ) n^{a}_{\m}\dot{\eta}^{b}_{0} &=& 0,
\label{normalitycond}\\
g_{ab}(\eta_0 ) n^{a}_{\m} n^{b}_{\n} &=& \d_{\m \n}.
\label{ortonormalitycond}
\eea
A escolha do conjunto $\{ n^{a}_{\m} \}$ n\~ao \'e \'unica, h\'a um grupo de calibre local $SO(D-1)$ correspondente as rota\cc\~oes do conjunto. Esta simetria de calibre \'e fixada impondo que os vetores normais sejam covariantemente constante
\be
\dot{\eta}^{a}_{0} \nabla_a n^{b}_{\m} = 0.
\label{covconst}
\ee
Neste calibre, as rela\cc\~oes entre os vetores normais tomam uma forma mais simples. Em particular, os vetores deste conjunto satisfazem a sequinte rela\cc\~ao de completeza
\be
g^{ab} = -\frac{1}{m^2}\dot{\eta}^{a}_{0}\dot{\eta}^{b}_{0} + n^{a}_{\m}n^{b}_{\n}\d^{\m\n}.
\label{completenessrel}
\ee
Considerando a primeira ordem $\eta^{a}_{1}(\t,\s)$ e admitindo perturba\cc\~oes 
co-moventes 
\be
\eta^{a}_{1}(\t , \s ) = \d x^{\m}(\t ,\s )n^{a}_{\m}.
\label{firstordperturb}
\ee 
As perturba\cc\~oes comoventes satisfazem as equa\cc\~oes de movimento com buraco negro de momento angular zero e tem como solu\cc\~ao geral uma corda bos\^onica fechada
\bea
\delta x^{\m} (\t , \s ) &=& \sum_{n \neq 0 } 
\sqrt{\frac{2|n|\O_{n}}{\a ' }}
\left[ \a^{\m}_{n} e^{-in(\O_n \t - \s)} + \b^{\m }_{n}e^{-in(\O_n \t + \s)} \right]
\nonumber\\
&+& \sqrt{\frac{l}{2m}} \left[ \a^{\m}_{0} e^{-i\frac{m\a '}{l}\t} + 
\b^{\m }_{0}e^{+i\frac{m\a '}{l}\t} \right].
\label{Fourierexp}
\eea
As frequ\^encias dos osciladores em unidades $ \hbar = 1$ s\~ao 
\be
\o_0 = m\a ' / l~~,~~ \o_n = \o_{-n} = |n| \O_n
\label{freqosc}
\ee
onde $ n = \pm 1, \pm 2, \ldots$. A frequ\^encia $\O_n$ \'e
\be
\O_n = \sqrt{1 + \frac{m^2 {\a '}^2}{n^2 l^2}}.
\label{Omega}
\ee
A solu\cc\~ao (\ref{Fourierexp}) satisfaz a equa\cc\~ao de movimento derivada da a\cc\~ao
\be
S_{2} = - \frac{1}{2\pi\a '} \int d\s d\t \sum_{\m =1}^{D-1} \left( \eta^{\a\b}
\6_{\a} \delta x^{\m} \6_{\b} \delta x^{\n} + \frac{m^2 {\a '}^2}{l^2} \d x^{\m} \d x_{\m}
\right). 
\label{truncatedaction}
\ee

\section{Quantiza\cc\~ao da Corda Bos\^onica no Espa\c{c}o-Tempo AdS}

A quantiza\cc\~ao da corda bos\^onica fechada no espa\cc o conforme AdS com $D =2+1$ e buracos negros est\'aticos se faz de maneira semelhante a quantiza\cc\~ao realizada no espa\cc o-tempo de Minkowski \cite{bhtz,bhtz1}. Ainda considerando perturba\cc\~oes em primeira ordem, podemos extender os resultados para espa\cc os AdS de dimens\~ao arbitr\'aria \cite{vesan}.

Os comutadores dos osciladores livres qu\^anticos satisfazem as rela\cc\~oes
\be
[\a^{\m}_{m}, \a^{\dagger\n}_{n} ] = [\b^{\m}_{m}, \b^{\dagger\n}_{n} ] = \d^{\m\n}\d_{mn}
~~,~~
[\a^{\m}_{m}, \b^{\n}_{n} ] = 0
~~,~~
[\a^{\m}_{0}, \a^{\dagger\n}_{0} ]  =  \d^{\m\n},
\label{osccommrel}
\ee
onde 
$\a^{\m}_{-n} = \a^{\dagger\m}_{n}$, $\a^{\m}_{-n} = \a^{\dagger\m}_{n}$, 
and $\b^{\m}_{0} = \a^{\dagger\m}_{0}$. 
As componentes do tensor energia-momento conservadas s\~ao
\bea
T_{--} = \frac{1}{2 \pi }\sum\limits_{n} L^{-}_{n} e^{-in(\s - \t)},
\label{T--}\\
T_{++} = \frac{1}{2 \pi }\sum\limits_{n} L^{+}_{n} e^{-in(\s + \t)}.
\label{T++}
\eea  
onde $ T_{++} = T_{--}=0$. Escrevendo os operadores de Virasoro para o modo zero
obtemos
\bea
L^{-}_{0} & = & \pi \a ' \sum\limits_{n > 0}
\left[ \frac{(\o_n - n)^2}{2n \O_n } \b^{\dagger}_{n} \cdot \b^{\dagger}_{n} +
\frac{(\o_n + n)^2}{2n \O_n } \a^{\dagger}_{n} \cdot \a^{\dagger}_{n} \right]\nonumber\\
&+&
\frac{\pi m {\a '}^2}{2l} \a^{\dagger}_{0} \cdot \a_{0} - 
\frac{\pi m^2 {\a '}^2}{2},
\label{L-}\\
L^{+}_{0} & = & \pi \a ' \sum\limits_{n > 0}
\left[ \frac{(\o_n + n)^2}{2n \O_n } \b^{\dagger}_{n} \cdot \b^{\dagger}_{n} +
\frac{(\o_n - n)^2}{2n \O_n } \a^{\dagger}_{n} \cdot \a^{\dagger}_{n} \right]\nonumber\\
&+&
\frac{\pi m {\a '}^2}{2l} \a^{\dagger}_{0} \cdot \a_{0} - 
\frac{\pi m^2 {\a '}^2}{2},
\label{L+}
\eea
onde $\cdot$ representa a soma em $\m = 1, 2, \ldots, D-1$. 

S\~ao estados f\'{\i}sicos os que satisfazem os v\'inculos dos operadores de Virasoro de modo zero
\be
\left( L^{-}_0 - 2 \pi \a ' \right) \left| \Psi_{phys} \right\rangle = 0
~~,~~
\left( L^{+}_0 - 2 \pi \a ' \right) \left| \Psi_{phys} \right\rangle = 0,
\label{constr}
\ee
e das simetrias na folha mundo: $\s \rightarrow \s + \xi$ and $\t \rightarrow \t + \zeta$ geradas por $P = L^{+}_{0} - L^{-}_{0}$ e $H = L^{+}_{0} + L^{-}_{0}$, respectivamente. Em $D \neq 2+1$, estes operadores n\~ao s\~ao mais os geradores da simetria conforme. Contudo, em primeira ordem, eles geram os mesmos v\'inculos acima.

O operador Hamiltoniano total e o operador de momento linear s\~ao
\bea
H & = & 2\pi\a ' \sum\limits_{n \geq 1}
\left( \frac{\O^{2}_{n} + 1}{\O_n} \right) \left( N_n + \overline{N}_{n}\right)
+\frac{\pi m {\a '}^2}{l} \a^{\dagger}_{0} \cdot \a_{0} - 
\pi m^2{\a '}^2,
\label{hamiltonian}\\
P & = & 4\pi\a ' \sum\limits_{n \geq 1}\left( N_n - \overline{N}_{n}\right).
\label{momentum}
\eea
onde 
\be
N_n  =  \frac{n}{2}\sum\limits^{D-1}_{\m = 1}\a^{\dagger \m}_{n} \a^{\m}_{n}
~~,~~
\overline{N}_n  =  \frac{n}{2}\sum\limits^{D-1}_{\m = 1}\b^{\dagger \m}_{n} \b^{\m}_{n},
\label{numberNbarN}
\ee
O momento linear para a corda fechada imp\~oe como v\' inculos para os estados f\'isicos
em $D =2 + 1$ 
\be
4\pi\a ' \sum\limits_{n \geq 1}\left( N_n - \overline{N}_{n}\right)
\left| \Psi_{phys} \right \rangle = 0.
\label{levelmatching}
\ee

No espa\c{c}o AdS de dimens\~{a}o arbitr\'{a}ria a unitariedade da teoria
\'{e} obtida impondo restri\c{c}\~{o}es de spin sobre as
representa\c{c}\~{o}es da \'{a}lgebra de Virasoro permitidas al\'{e}m dos
v\'{\i}nculos de Virasoro porque estes n\~{a}o eliminam completamente os
estados de norma negativa. As f\'{o}rmulas obtidas acima para o buraco negro
AdS em $D=2+1$ n\~{a}o dependem da massa do buraco negro e tamb\'{e}m, essa
configura\c{c}\~{a}o de fundo \'{e} assitoticamente AdS. Consequentemente,
podemos generalizar para espa\c{c}o AdS de dimens\~{a}o arbitr\'{a}ria. \'{E}
importante observar que $L_{n}^{+}$ e $L_{n}^{-}$ n\~{a}o geram simetrias
exatas no espa\c{c}o AdS de dimens\~{a}o arbitr\'{a}ria. Entretanto, o
Hamiltoniano correspondente e a condi\c{c}\~{a}o operatorial de n\'{\i}veis
iguais (\ref{levelmatching}) s\~{a}o obtidas quando da generaliza\c{c}\~{a}o da dimens\~{a}o do
espa\c{c}o AdS em primeira ordem.

Ainda observamos que as excita\c{c}\~{o}es da corda no espa\c{c}o AdS oscilam
no tempo. Apesar de que poss\'{\i}veis instabilidades n\~{a}o se desenvolvem
devido ao car\'{a}ter n\~{a}o negativo da gravidade local.  

\newpage

\chapter{Din\^amica de Campos T\'ermicos}

Diversos formalismos introduzem a temperatura em teoria de campos. Os sistemas
f\'{\i}sicos encontrados na natureza, geralmente n\~{a}o est\~{a}o
completamente isolados; \'{e} necess\'{a}ria a descri\c{c}\~{a}o de sistemas
com infinitos graus de liberdade \`{a} temperatura finita. Como alternativa ao
m\'{e}todo de Matsubara \cite{mt}, Takahashi e Umezawa \cite{tu1,tu2} adotaram um
postulado fundamental, construiram estados de v\'{a}cuo t\'{e}rmico dependente
da temperatura que se relacionam, via transforma\c{c}\~{a}o de Bogoliubov. Foi
poss\'{\i}vel construir todos estados t\'{e}rmicos, estabelecendo-se a DCT.

\section{Postulado Fundamental da DCT}

O formalismo desenvolvido por Umezawa e Takahashi \cite{tu1,tu2}, inspirado em \cite{mt} \'e baseado no c\'alculo de m\'edias estat\'isticas de uma vari\'avel din\^amica $A$ com valor esperado deste operador num v\' acuo dependente da temperatura (v\'acuo t\'ermico). Como postulado
\be
\la A \ra=Z^{-1}(\b_T)tr\;[e^{-\b_T {\cal{H}}}A]=\la 0(\b_T)|A|0(\b_T) \ra \label{0},
\ee
onde ${\cal{H}}=H-\mu N$, $Z(\b_T)=tr\;[e^{-\b_T {\cal{H}}}]$ e $\b=\frac {1}{{k}_{B}T}$ sendo $H$ a Hamiltoniana total, $\mu$ o potencial qu\'{\i}mico, $N$ o n\'umero de part\'{\i}culas e ${k}_B$ a constante de Boltzmann. 

\section{DCT no Formalismo Can\^onico}

Considerada a m\'edia estat\'{i}stica
\be
\la 0(\b_T)|A|0(\b_T) \ra=Z^{-1}(\b_T) \sum_{n}\la n|A|n \ra e^{-\b_T {\o}_{n}},   \label{1111}
\ee
a expan\cc\~ao do v\'acuo em termos de uma base $\{ |n \ra \}$ do espa\c{c}o de Hilbert \'e dada pela express\~ao
\be
|0(\b_T) \ra=\sum _{n}|n \ra \la n |0(\b_T) \ra=\sum _{n}f_{n}(\b_T)|n \ra ,    \label{2a}
\ee
onde $f_{n}(\b_T)$ s\~ao coeficientes a serem determinados. Da ortogonalidade entre os estados da base $\{ |n \ra \}$ 
e a normaliza\cc\~ao do estado $|0(\b_T) \ra$ resulta
\be
f^{\ast}_{n}(\b_T)f_{m}(\b_T) = Z^{-1}(\b_T)e^{-\b_T \o_{n}} \d_{nm}. \label{3a}
\ee
Observamos que a express\~{a}o (\ref{3a}) \'{e} correta somente se $\{f_{n}(\beta_{T})\}$
s\~{a}o coeficientes vetoriais.\ O estado de v\'{a}cuo t\'{e}rmico devera ser
expandido por $\{\left\vert n\right\rangle \}$ e $\{f_{n}(\beta_{T})\}$.
Existe necessidade de dobrar os graus de liberdade do sistema e utilizar para
isto um espa\c{c}o auxiliar n\~{a}o f\'{\i}sico $\tilde{\HH}$, id\^{e}ntico
e ortogonal ao espa\c{c}o f\'{\i}sico inicial $\HH$; para este espa\c{c}o
extendido $\hat{\HH}$ por constru\c{c}\~{a}o
\be
\hat{\HH} = \HH \otimes \tilde{\HH}.
\label{totalHilbert}
\ee
Se $\{ |n \ra \}$ s\~ao autoestados do Hamiltoniano $H$ e $\{ \tilde{|n \ra} \}$ autoestados da sua c\'opia $\tilde{H}$ obedecem as rela\cc\~oes
\be
H |n \ra = \o_n |n \ra ~~,~~ \tilde{H} \tilde{|n \ra} = \o_n \tilde{|n \ra},
\label{copiesH}
\ee
onde $\la n | m \ra = \la \tilde{n}|\tilde{m} \ra = \d_{nm}$ e $\o_n$ \'e a mesma frequ\^encia do sistema f\'isico. Um vetor de estado do sistema total do $\hat{\HH}$ \'e construido como
\be
|n,\tilde{n}\ra = |n\ra \otimes |\tilde{n} \ra ,
\label{vettotal}
\ee
o que determina o seguinte coeficiente vetorial
\be
f_{n}(\b_T)=e^{-\b \o_{n}/2}Z^{-1/2}(\b_T)| \widetilde n \ra.   \label{def}
\ee
Finalmente, o estado de v\'acuo t\'ermico \'e
\be
|0(\b_T) \ra = \sum_{n}e^{-\b_T \o_{n}/2}Z^{-1/2}(\b_T)|n,\widetilde n \ra. \label{vab}
\ee
Vamos representar o estado de v\'acuo \`{a} temperatura zero como $ | 0 \ra \! \ra = |0,\tilde{0}\ra$. Com a condi\cc\~ao de que $ | 0 \ra = |0,\tilde{0}\ra$ \'e normalizado $\la 0 | 0 \ra = 1$ obtemos a fun\cc\~ao de parti\cc\~ao
\be
Z(\b_T)=\sum_{n}e^{-\b_T \o_{n}} \langle n|n\rangle=tr\;[e^{-\b {\cal{H}}}].
\label{partfuncttfd}
\ee
O valor m\' edio do operador $A$ no estado de v\'acuo t\'ermico, est\'a de acordo com o postulado fundamental da DCT.

Considerando um sistema bos\^onico, os operadores $A$ e $\tilde{A}$ que atuam nos espa\cc os $\HH$ e $\tilde{\HH}$, respectivamente, comutam entre si
\be
[ A, \tilde{A} ] = 0.
\label{commutatorHH}
\ee
A t\'itulo de exemplo, consideramos o Hamiltoniano de um oscilador bos\^onico \`{a} temperatura zero
\be
{\bo{H}}=\o a^{\dagger}a.
\ee
Tem-se as rela\cc\~oes de comuta\cc\~ao
\be
[a,a^{\dagger}]=1 \ \ \ \ \mbox{e} \ \ \ \ [a,a]=[a^{\dagger},a^{\dagger}]=0.  \label{com0}
\ee
Os estados no espa\cc o de Fock correspondem a
\be
|n\ra = \frac{(a^{\dagger})^n}{\sqrt{n!}}|0 \ra ~~,~~a|0\ra = 0.
\label{estadosFock}
\ee
Duplicando o sistema original, o Hamiltoniano $\tilde{\HH}$ no espa\cc o auxiliar \'e
\be
\widetilde{\bo {H}}=\o  \widetilde a^{\dagger}  \widetilde a,
\label{copia}
\ee
e as seguintes rela\cc\~oes de comuta\cc\~ao s\~ao satisfeitas
\be
[ \widetilde a,  \widetilde a^{\dagger}]=1 \ \ \ \ \ \mbox{e} \ \ \ \ \  [ \widetilde a, \widetilde a]=[ \widetilde a^{\dagger}, \widetilde a^{\dagger}]=0. 
\label{com1}
\ee
e
\be 
[a, \widetilde a]=[a^{\dagger}, \widetilde a^{\dagger}]=[a, \widetilde a^{\dagger}]=[a^{\dagger}, \widetilde a]=0.
\label{com2}
\ee
O v\'acuo do sistema duplicado deve satisfazer
\be
|0 \ra \otimes \w {|0 \ra} = |0\ra \! \ra.
\label{vactot}
\ee
O estado de v\'acuo t\'ermico para o sistema extendido \'e obtido via uma transforma\cc\~ao unit\'aria \cite{tu1,tu2}
\be
|0(\b_T)\ra =e^{-iG(\th)}|0\ra \! \ra ,  \label{tbv}
\ee
cujo gerador \'e o operador de Bogoliubov
\be
G(\th)=G(\th)^{\dagger}=-i\th(\b_T)( \widetilde{a}a- {a}^{\dagger}\widetilde{a}^{\dagger}),  \label{gtb}
\ee 
Escolhendo o par\^ametro $\th(\b_T) \in R$, $G_B = G_B^{\dagger}$ \'e hermitiano. 
Definindo 
\be
u(\b_T)=(1-e^{-\b \o})^{-\frac{1}{2}} = \cosh\th(\b_T) , \label{u11}
\ee
e
\be
\v(\b_T)=(e^{\b \o}-1)^{-\frac{1}{2}}=\sinh\th(\b_T) ,  \label{v11}
\ee
onde $f_{B}$ \'e a distribui\c{c}\~ao de Bose, o estado de v\'acuo t\'ermico para o sistema total resulta
\be
|0(\b_T)\ra = \frac{1}{\cosh \th(\b_T)} \exp \left[ \tanh (\th (\b_T ))\right]|0\ra \! \ra .
\label{vactotal}
\ee
A transforma\cc\~ao de Bogoliubov atuando sobre os operadores de aniquila\cc\~ao $a$ e $\tilde{a}$ em $T=0$ mapea em $a(\b_T)$ e $\tilde{a}(\b_T)$, respectivamente
\be
a(\b_T)= e^{-iG}a e^{iG}, 
\label{trBog}
\ee
devido a unitariedade da transforma\cc\~ao gerada pelo operador de Bogoliubov. A transforma\cc\~ao (\ref{trBog}) pode ser escrita como uma transforma\cc\~ao linear
\bea
a &=& u(\b_T)a(\b_T)+\v(\b_T) \widetilde{a}^{\dagger}(\b_T)\ ,\ a^{\dagger}=u(\b_T)a^{\dagger}(\b_T)+\v(\b_T) \widetilde{a}(\b_T) \label{lin1}\\
\widetilde{a} &= & u(\b_T) \widetilde{a}(\b_T)+\v(\b_T){a}^{\dagger}(\b_T) \ , \
 \widetilde{a}^{\dagger}=u(\b_T) \widetilde{a}^{\dagger}(\b_T)+\v(\b_T){a}(\b_T).
\label{lin2}
\eea
Como esperado para o v\'acuo t\'ermico
\bea
a(\b_T) |0(\b_T ) \ra \! \ra &=& \tilde{a}(\b_T)|0(\b_T ) \ra \! \ra =0, \nonumber\\
\la \! \la 0(\b_T ) | a(\b_T)^{\dagger}  &=& \la \! \la 0(\b_T ) | \tilde{a}(\b_T)^{\dagger} = 0.
\label{termvacops}
\eea
Com uma sequ\^encia de atua\cc\~oes dos operadores  ${a(\b_T)}^{\dagger}$ e ${\tilde{a}(\b_T)}^{\dagger}$, obtemos os estado t\'ermicos de Fock
\be
|0(\b_T)\ra \! \ra , {a(\b_T)}^{\dagger}|0(\b_T)\ra \! \ra, {\tilde{a}(\b_T)}^{\dagger}|0(\b_T)\ra \! \ra , \ldots , 
\frac{1}{\sqrt{n!}\sqrt{m!}}({a(\b_T)}^{\dagger})^{n} ({\tilde{a}(\b_T)}^{\dagger})^{m}|0(\b_T)\ra \! \ra .
\label{esttermFoc}
\ee
As rela\cc\~oes de comuta\cc\~ao entre os operadores t\'ermicos, i. e. operadores com depend\^encia em $\b_T$ s\~ao as mesmas que as apresentadas no sistema extendido em $T=0$. Explorando os comutadores de $G$ com os operadores dos osciladores 
\be
[G,a] =  -i\th (\b_T) \widetilde{a}^{\dagger}~,~
[G, \widetilde{a}] =-i\th (\b_T){a}^{\dagger}~,~ 
[G,a^{\dagger}]=-i\th (\b_T) \widetilde{a}~,~
[G, \widetilde{a}^{\dagger}]=-i\th (\b_T)a
\label{comBog}
\ee
onde, por simplicidade $\th = \th(\b_T)$ \'e sempre dependente da temperatura. Resulta que o gerador da transforma\cc\~ao $G$ \'e conservado (can\^onico):
\be 
i\dot{G} = [G,H]=0.         \label{gerador}
\ee
Observamos que qualquer estado de ocupa\cc\~ao pode ser obtido.

\section{Formalismo para Campos Livres e a Entropia}

Considerando o sistema total, a Lagrangeana $\hat{L} = L - \tilde{L}$ leva a escrever a Hamiltoniana extendida como $\hat{H} = H - \tilde{H}$. A Hamiltoniana $\hat{H}$ \'e invariante por transforma\cc\~ao de Bogoliubov. Considera-se o volume finito onde se quantizam os campos que expandidos em ondas planas dependem dos operadores ${a}_{\ve{k}}(\b)$, 
$\widetilde{a}_{\ve{k}}(\b)$ e os operadores til conjugados. A transforma\cc\~ao de Bogoliubov \'e unit\'aria $U = \exp(-iG)$ onde o gerador de Bogoliubov para o campo \'e 
\be
G=-i \sum_{\ve{k}} \th_{\ve{k}}(a^{\a}_{\ve{k}}  \widetilde{a}^{\a}_{\ve{\k}}-\widetilde{a}_{\ve{\k}}a_{\ve{k}}),  \label{opbogcamplivre}
\ee
e satisfaz a rela\cc\~ao de comuta\cc\~ao $[G,\hat{H}]=0$. Os operadores de aniquila\cc\~ao dependentes de temperatura s\~ao obtidos dos operadores de aniquila\cc\~ao \`{a} temperatura zero da seguinte form 
\be
{a}_{\ve{k}}(\b)=e^{-iG}a_{\ve{k}}e^{iG}={a}_{\ve{k}}\cosh{\th_{\ve{k}}}(\b)- \widetilde{a}^{\a}_{\ve{k}}\sinh{\th_{\ve{k}}}(\b), \label{oaclivre}
\ee
\be
\widetilde{a}_{\ve{k}}(\b)=e^{-iG} \widetilde{a}_{\ve{k}}e^{iG}= \widetilde{a}_{\ve{k}}\cosh {\th_{\ve{k}}}(\b)-{a}^{\a}_{\ve{k}}\sinh {\th_{\ve{k}}}(\b)\ . \label{oatclivre}
\ee
Explorando a liberdade que a transforma\cc\~ao de Bogoliubov oferece, podemos definir o v\'acuo t\'ermico
\be
|0(\b)\ra =U(\th)|0(\b)\ra= e^{-iG}|0\ra \! \ra . 
\label{vactermclivre}
\ee
que deve satisfazer as sequintes rela\cc\~oes
\be
a_{\ve{k}}(\b)|0(\b)\ra =e^{-iG}a_{\ve{k}}e^{iG}e^{-iG}|0\ra \! \ra =e^{-iG}a_{\ve{k}}|0\ra \! \ra =0, \label{vaclivre}
\ee
\be
 \widetilde{a}_{\ve{k}}(\b)|0(\b)\ra =e^{-iG} \widetilde{a}_{\ve{k}}e^{iG}e^{-iG}|0\ra \! \ra .
=e^{-iG} \widetilde{a}_{\ve{k}}|0\ra \! \ra 
=0.  \label{vatclivre}
\ee

Para o sistema de osciladores que comp\~oem o campo podemos definir fun\cc\~oes termodin\^amicas.
As grandezas entropia e energia livre de Helmholtz t\^em papel importante no formalismo DCT. O operador $K$ definido como
\be
K=-\sum_{\ve{k}} \le(a^{\dagger}_{\ve{k}}a_{\ve{k}}\ln{\sinh^{2}{\th_{\ve{k}}(\b_T)}}-a_{\ve{k}}a^{\dagger}_{\ve{k}}\ln{\cosh^{2}{\th_{\ve{k}}(\b_T)}}\ri ).  
\label{centropia}
\ee
O operador $\tilde{K}$ \'e obtido por conjuga\cc\~ao til do operador $K$. Pode-se mostrar que o operador $\hat{K}= K- \widetilde{K}$ satisfaz as rela\c{c}\~oes
\be 
(K- \widetilde{K})|0(\b)\ra = 0 ~~,~~ [K- \widetilde{K},G] = 0. \label{Kprop}
\ee
com $G$ dado pela equa\c{c}\~ao~(\ref{opbogcamplivre}). 

Para o caso bos\^onico, usando as rela\cc\~oes de comuta\cc\~ao reescrevemos o estado de v\'acuo t\'ermico \cite{tu1,tu2}
\be
|0(\b_T)\ra =e^{-K/2} \le \{\exp{\sum_{\ve{k}}a^{\dagger}_{\ve{k}} \widetilde{a}^{\dagger}_{\ve{k}}}\ri \}|0\ra \! \ra . \label{vtk}
\ee
A entropia no sistema Gr\~a-Can\^onico \cite {greiner} \'e calculada como o valor esperado m\'edio do operador $K$ no v\'acuo t\'ermico 
\be
S=k_{B} \langle K\ra =k_{B} \langle 0(\b_T)|K|0(\b_T)\ra . 
\label{entropiaK}
\ee
Um c\'alculo simples leva a seguinte express\~ao para a entropia 
\be
S=k_{B}\sum_{{k}}\le \{ (1+ \langle n_{{k}}\ra )\ln(1+ \langle n_{{k}}\ra )- \langle n_{{k}}\ra \ln  \langle n_{{k}}\rangle \ri \}
\label{entropiaS}
\ee
onde $n_{k}$ representa o n\'umero m\'edio de ocupa\c{c}\~ao do estado $k$.

Usando a formula\cc\~ao can\^onica do formalismo DCT, pretendemos obter os estados \`a temperatura finita para a corda bos\^onica com v\'arias condi\cc\~oes de contorno como descrito no capitulo anterior, assim como o operador entropia e a energia livre de Helmholtz calculada a partir da sua defini\cc\~ao
\be
F = -TS+ \langle H\ra -\mu \langle N\rangle .
\label{energialivre}
\ee

\section{Axiomas da DCT}

Para quaisquer operadores $A$ e $\tilde{A}$ por atuarem respectivamente no espa\cc o $\HH$ e em espa\cc o auxiliar fict\'icio $\tilde{\HH}$ ortogonal a $\HH$, temos que o comutador $[A,\tilde{A]}=0$. Exis\-tem, al\'em disso, um mapeamento entre o conjunto de operadores $\{ A \}$ e $\{ \tilde{A} \}$ que obedece as denominadas regras de conjuga\cc\~ao til. A temperatura entra na teoria atrav\'es de condi\cc\~oes que relacionam a forma na qual $A$ e $\tilde{A}$ atuam no v\'acuo t\'ermico $|0(\b_T )\ra \ra$. Esta \'e a condi\cc\~ao de estado t\'ermico,tamb\'em denominada de {\em regra de substitui\cc\~ao til}. Uma teoria DCT para a teoria qu\^antica de campos (TQC), pode ser melhor construida a partir de axiomas b\'asicos da DCT \cite{tu1,tu2,ubook}. 

Vamos enunciar os axiomas, considerando dois conjuntos de operadores
$\Im=\{A\}$  e $\w{\Im}=\{\w{A}\}$, ent\~ao

{\it Axioma 1 }. A tempos iguais, vari\'aveis din\^amicas pertencentes a diferentes sub-espa\cc os
($A \in \Im \ \mbox{e} \ \w{B} \in \w{\Im} $) s\~ao independentes, ou seja
\be
[A,\w{B}]=0.
\label{axi1}
\ee

{\it Axioma 2 }. Existe um mapeamento um a um entre os espa\cc os ortogonais de\-no\-minado de {\em conjuga\cc\~ao til}; 
para quais $A$ e $B$ $\in$  $\Im$ e $\tilde{A}$ e $\tilde{B}$ $\in$ $\w{\Im}$ e $c_1, c_2$ dois n\'umeros complexos, valem as regras de conjuga\cc\~ao til:
\bea
& & \ \w{(AB)}=\w{A}\w{B},  \\ 
& & \ \w{{(c_{1}A+c_{2}B)}}=c^{\ast}_{1}\w{A}+c^{\ast}_{2}\w{B}, \\
& & \ \w{A^{\dagger}}=\w{A}^{\dagger}.
\label{axi2}
\eea

{\it Axioma 3 }. O v\'acuo t\'ermico \'e invariante sob as regras de conjuga\cc\~ao til 
\be
\w{|0(\b_T) \ra}=|0(\b_T) \ra.
\label{axi3}
\ee

{\it Axioma 4 }. Transla\cc\~ao espa\cc o-temporais s\~ao induzidas pelo operador energia-momento $P_\m \in \Im$ da
\be
A(x)=e^{iP_{\mu}x^{\mu}}Ae^{-iP_{\mu}x^{\mu}}.
\label{axi4}
\ee
{\it Axioma 5 }. O v\'acuo t\'ermico \'e definido pelas rela\cc\~oes operatoriais chamadas de {\em condi\cc\~oes de estado t\'ermico} 
\be
A(t,\ve{x})|0(\b_T) \ra=\s \w{A}^{\dagger}(t-i\b/2, \ve{x})|0(\b_T) \ra, 
\label{axi5a}
\ee 
\be
\la O(\b_T)|A(t,\ve{x})=\la O(\b_T)| \w{A}^{\dagger}(t+i\b/2), \ve{x}) \s^{\ast}, 
\label{axi5b}
\ee 
Se $A$ \'e uma vari\'avel bos\^onica, escolhemos $\s = 1$. 

{\it Axioma 6 }. A dupla conjuga\cc\~ao til \'e definida como
\be
\w{\w{A}}= \s A.
\label{axi6}
\ee
onde $\s =1$ para bosons e $\s = -1$ para f\'ermions.

A import\^{a}ncia das regras de conjuga\c{c}\~{a}o til \'{e} a de que todas as
rela\c{c}\~{o}es usuais da TQC, por exemplo rela\c{c}\~{o}es de
comuta\c{c}\~{a}o, e equa\c{c}\~{o}es de Heisenberg podem ser generalizadas
para DCT. A condi\c{c}\~{a}o de estado t\'{e}rmico, al\'{e}m de fundamental
para definir o v\'{a}cuo t\'{e}rmico, mostra que existe sempre uma
combina\c{c}\~{a}o de operadores $A(x)$ e $A^\dag(x)$ que aniquila o v\'{a}cuo
t\'{e}rmico. Esta caracteristica usualmente n\~{a}o existe em TQC. Podemos
generalizar o Axioma 1. Sejam $A(x)$ e $\widetilde{B}(y)$ ent\~{a}o eles
comutam em todo espa\c{c}o-tempo
\be 
[A(x),\w{B}(y)]=0.
\ee
Se realizarmos uma opera\cc\~ao $\dagger$ e uma $\tilde{}$ ou uma opera\cc\~ao $\tilde{}$ e uma $\dagger$, pelo Axioma 2 verificamos que os coeficientes dos operadores permanecem inalterados. Podemos considerar um axioma suplementar a constru\cc\~ao da Lagrangeana e Hamiltoniana extendidas
\be
\hat{H}=\sum_{\a}\e^{\a}H^{\a}=H-\w{H}, \ \  \ \hat{L}=\sum_{\a}\e^{\a}L^{\a}=L-\w{L}.
\ee
Decorre dos axiomas a seguinte propriedade do v\'acuo t\'ermico
\be
a(\b_T,t)|0(\b_T)\ra= \w{a}(\b_T,t)|0(\b_T)\ra=\la 0(\b_T)|a^{\a}(\b_T,t)=\la 0(\b_T)|\w{a}^{\a}(\b_T,t)=0. \label{oavt}
\ee 
A rela\cc\~ao entre a conjuga\cc\~ao til e a conjuga\cc\~ao hermitiana do operador $A(t)$ num instante $t$ \'e
\be
A(t)|0(\b_T)\ra= {\w{A}}^{\dagger}(t-i\b/2)|0(\b_T)\ra,
\ee
\be
A^{\dagger}(t)|0(\b_T)\ra={\w{A}}(t-i\b/2)|0(\b_T)\ra,
\ee
onde 
\be
a(\b_T,t)=f^{1/2}(-i\6_{t}) \left (A(t+i\b/2)-{\w{A}}^{\dagger}(t) \ri ),
\ee 
\be
\w{a}(\b_T,t)=f^{1/2}(-i\6_{t})^{\ast} \left (\w{A}(t-i\b/2)-A^{\dagger}(t) \ri),
\ee
e
\be
f(\o)=\frac{1}{e^{\b \o} - 1}, \label{f}
\ee
\'e a fun\cc\~ao de Bose-Einstein. A rela\cc\~ao entre os operadores \`a temperatura nula e os operadores \`a temperatura finita pode ser derivada a partir dos axiomas do formalismo DCT
\be
A=e^{iG(\b_T)}a(\b_T,t)e^{-iG(\b_T)},
\ee
e tamb\'em, a rela\cc\~ao entre o v\'acuo t\'ermico e o v\'acuo duplicado \`a temperatura nula 
\be
|0(\b_T)\ra=e^{-iG(\b_T)}|0,\w{0}\ra, \label{mv}
\ee
com o operador de Bogoliubov definido pela 
\be
G(\b_T)=G^{\dagger}(\b_T)=-\w{G}(\b_T).
\ee

\newpage

\chapter{Estados da Corda Bos\^onica T\'ermica no Formalismo DCT}

\qquad Neste cap\'{\i}tulo, construimos os estados de corda bos\^{o}nica
t\'{e}rmica aberta no espa\c{c}o de Minkowski e calculamos a entropia desses
estados \cite{eg1}. Na sequ\^encia, co\-nstruimos os estados de corda bos\^{o}nica fechada t\'{e}rmica 
no espa\c{c}o AdS em primeira aproxima\cc\~ao; calculamos a entropia usando o formalismo DCT \cite{eg2} 
e discutimos a rela\c{c}\~{a}o entre a
Hamiltoniana no espa\c{c}o de Hilbert total e o espa\c{c}o de Hilbert f\'{\i}sico \cite{eg3}. Estas contribui\c{c}\~{o}es e possibilidades abertas ser\~{a}o comentadas no final
do cap\'{\i}tulo.

\section{Estados da Corda Aberta T\'ermica no Espa\cc o-Tempo de Minkowski}

Inicialmente para construir os estados da corda, escrevemos os operadores de cria\c{c}\~{a}o e aniquila\c{c}\~{a}o dos osciladores da corda f\'{\i}sica obtidos no primeiro cap\'itulo
\be
A^{\m}_n = \frac{1}{\sqrt{n}}\a^{\m}_n ~~~;~~~A^{\m \dagger}_n = 
\frac{1}{\sqrt{n}}\a^{\m}_{-n}~~,
\label{leftca1}
\ee
C\'{o}pias id\^{e}nticas de operadores s\~{a}o escritas, para atua\c{c}\~{a}o no espa\c{c}o aix\'{\i}liar $\hat{H}$
\be
\w{A}^{\m}_n = \frac{1}{\sqrt{n}}\w{\a}^{\m}_n ~~~;~~~\w{A}^{\m \dagger}_n = 
\frac{1}{\sqrt{n}}\w{\a}^{\m}_{-n}~~,
\label{leftca1tilde}
\ee
Os operadores satisfazem a \'algebra
\be
[A^{\m}_n,A^{\nu \dagger}_m]=[\w{A}^{\m}_n,\w{A}^{\nu \dagger}_m]=\d_{n+m} \h^{\mu \nu} \ \ , \ \ 
[A^{\m}_{n},{\w{A}}^{\nu}_{m}]=[A^{\m}_n,{\w{A}}^{\nu \dagger}_m]=0.\label{comutad}
\ee
O espa\cc o de Fock do sistema total \'e o produto tensorial dos espa\cc os de Fock de cada corda.
Considerando $T=0$, o estado de v\'{a}cuo dos osciladores da corda \'{e}
\be
\left| 0 \right\rangle  ~= ~\left| 0 \right\rangle  \left| p \right\rangle ,
\label{vaccbosTzero}
\ee 
onde para obtermos o estado fundamental de v\'{a}cuo devemos considerar a parte de
momento do centro de massa. Assim,
\be
|0 \ra \! \ra \otimes |p \ra \otimes | \tilde{p} \ra = |0,\tilde{0}\ra  |p , \tilde{p} \ra
\label{estvac}
\ee
representa o v\'{a}cuo fundamental em $T=0$ e
\bea
A^{\m}_n \left| 0 \right\rangle ~ &=& ~ 0 ~~,~~~ \forall n, \label{vacosc}\\
\hat{p}^{\m} \left| p \right\rangle ~ & = & ~ p^{\m} \left| p \right\rangle 
\label{vacmom1}.
\eea
Uma vez duplicado o n\'{u}mero de
graus de liberdade; faz-se uso dos operadores unit\'{a}rios de Bogoliubov
$G_{n}^{\mu}$, para obtermos a descri\c{c}\~{a}o t\'{e}rmica. Definimos
\be
G^{\m}_n ~=~-i \th_n(\b_T)(A_n \cdot \tilde{A}_n - \tilde{A}_n^{\dagger} \cdot
 A_n^{\dagger} ).
\label{bogolopcb1}
\ee  
onde $\theta(\beta_{T})$ \'{e} um par\^{a}metro real que depende da
estat\'{\i}stica do n - \'{e}simo modo $\cosh\theta_{n}(\beta_{T}%
)=(1-e^{\beta_{T^{M}}})^{-1}$. $A_{n} \cdot \tilde{A}_{n}$ representa o produto escalar
$A_{n}^{\mu} \widetilde{A}_{\mu\text{ }n}$ no espa\c{c}o de Minkowski.
Os operadores $G_{n}^{\mu}$ s\~{a}o hermitianos e $G_{\left\vert n\right\vert
}=-G_{-n}$ para $n<0$. Escolhido o calibre de cone de luz, onde
$x^{\circ}\pm x^{25}$; $\mu=1,\cdots,24;$ $D=26$\bigskip, sem anomalias, sem
estados fantasmas $G_{n}=\sum_{\mu=1}^{24}G_{n}^{\mu}$.

As rela\c{c}oes de comuta\c{c}\~{a}o entre operadores de Bogoliubov e os osciladores s\~{a}o:
\be
[G_{n},A^{\mu}_{n}]=-i \th_{n}({\b}_{T}){\w{A}}_{n}^{\mu \dagger}, \ \ \ \  
[G_{n},A^{\mu \dagger}_{n}]=-i \th_{n}({\b}_{T}){\w{A}}_{n}^{\mu}, 
\label{commbogop1}
\ee
\be
[G_{n},{\w{A}}^{\mu}_{n}]=-i \th_{n}({\b}_{T}){A}_{n}^{\mu \dagger}~~,~~   
[G_{n},{\w{A}}^{\mu \dagger}_{n}]=-i \th_{n}({\b}_{T}){A}_{n}^{\mu}, 
\label{commbogca2}
\ee
Vamos construir o estado de v\'{a}cuo t\'{e}rmico e os operadores de
aniquila\c{c}\~{a}o e cria\c{c}\~{a}o t\'{e}rmicos para corda aberta
\be
\left. \left| 0(\b_T ) \right\rangle \! \right\rangle_{osc} ~= ~\prod_{ m > 0} 
e^{-iG_m} 
\left. \left| 0 \right\rangle \! \right\rangle ,
\label{vacT1}
\ee
onde $\left. \left| 0 \right\rangle \! \right\rangle= 
\left| 0 \right\rangle\tilde{\left| 0 \right\rangle} $. 
O v\'acuo t\'ermico do sistema total \`a temperatura finita cont\'em contribui\cc\~oes do momento linear 
\be
\left. \left| 0 (\b_T) \right\rangle \! \right\rangle ~=~ 
\left. \left| 0(\b_T) \right\rangle \! \right\rangle_{osc}
\left| p \right\rangle \left| \tilde{p} \right\rangle. 
\label{totvacT1}
\ee
Os operadores que aniquilam o v\'{a}cuo t\'{e}rmico s\~{a}o 
\be
A^{\m}_{n}(\b_T) ~= ~ e^{-iG_n}A ^{\m}_{n}e^{iG_n}~~~,~~~
\tilde{A}^{\m}_{n}(\b_T) ~= ~ e^{-iG_n}\tilde{A} ^{\m}_{n}e^{iG_n}.
\label{annihT1}
\ee
e os operadores que criam estados a partir do v\'{a}cuo t\'{e}rmico s\~ao conjugados hermitianos destes.
Os estados do sistema a temperatura finita s\~{a}o obtidos atuando no
v\'{a}cuo t\'{e}rmico com os operadores t\'{e}rmicos de cria\c{c}\~{a}o e
destrui\c{c}\~{a}o. Os estados obtidos pertencem a um espa\c{c}o de Fock
t\'{e}rmico. As coordenadas de momento e centro de massa \ da corda s\~{a}o
invariantes por transforma\c{c}\~{o}es de Bogoliubov, ou seja, todos os
operadores dos osciladores comutam com os operadores $x, \tilde{x}, p, \tilde{p}$ ,
podemos tratar a corda como um conjunto de osciladores bos\^{o}nicos. Os
operadores de entropia para a corda bos\^{o}nica aberta, diretamente de suas
defini\c{c}\~{o}es s\~{a}o
\be
K ~= ~\sum_{\m = 1}^{24}\sum_{n=1}^{\infty}( A^{\m \dagger}_n A^{\m}_n 
\log \sinh^2 \th_n -A^{\m}_n A^{\m \dagger}_n \log \cosh^2 \th_n )
\label{entropy1}
\ee
e o operador $\tilde{K}$ obtido atrav\'es da conjuga\cc\~ao til do $K$.
O v\'{a}cuo \'{e} invariante sob a opera\c{c}\~{a}o til, todas as
informa\c{c}\~{o}es est\'{a} contida em operadores sem til, assim os elementos
de matriz que interessam s\~{a}o os do operador K. Foi \'{u}til escrever
\be
K ~ =~ \sum_{\m = 1}^{24} K^{\m},
\label{decomentrop1}
\ee
A entropia da corda representa a soma das entropias de todos
osciladores em todas as dire\c{c}\~{o}es. Pretendemos encontrar a entropia da
corda associada mais geral da equa\c{c}\~{a}o de movimento.
A entropia \'{e} fun\c{c}\~{a}o do campo $x(\tau,\sigma)$ que descreve a
folha mundo e as condi\c{c}\~{o}es de contorno nas equa\c{c}\~{o}es de
movimento informam qual \'{e} a depend\^{e}ncia com os par\^{a}metros da folha
mundo. Vamos escolher a solu\c{c}\~{a}o geral com as c.c. $NN$ para
calcularmos os elementos de matriz, inicialmente.

Os elementos de matriz $<X^{\m}(\beta_{T})\left\vert K\right\vert
X^{\nu}(\beta_{T})>$ podem ser separados em duas partes: a primeira que
cont\'{e}m as coordenadas e momento do centro de massa \ e a parte que
cont\'{e}m a informa\c{c}\~{a}o \ dos osciladores, ou seja
\bea
\left\langle \!\left\langle X^{\mu}(\b_T)\left| K^{\rho} \right| 
X^{\m}(\b_T )\right\rangle\!\right\rangle
~&=&~{\mbox{termos do c. m.}} \nonumber\\
&-& 2\a ' \sum_{n,k,l >0}\f{e^{i(l-n)\t}}{\sqrt{ln}}\cos n\s \cos l\s
\left[ (T_1)^{\m\rho\n}_{nkl} + (T_2)^{\m\rho\n}_{nkl}\right],\nonumber\\
\label{matrentr1}
\eea
onde
\bea
(T_1)^{\m\rho\n}_{nkl} ~&=&~
\langle \tilde{0}, 1^{\m}_n |
\prod_{m>0}e^{-iG_m}A^{\rho \dagger}_{k}A^{\rho}_{k}\log \sinh^2 \theta_k 
\prod_{s>0}e^{iG_s}
| 1^{\n}_l , \tilde{0} \rangle
\langle \tilde{p}, p| \tilde{q}, q\rangle,
\nonumber\\
(T_2)^{\m\rho\n}_{nkl} ~&=&~
- \langle \tilde{0} , 1^{\m}_n |
\prod_{m>0}e^{-iG_m}A^{\rho }_{k}A^{\rho \dagger}_{k}\log \cosh^2 \theta_k 
\prod_{s>0}e^{iG_s}
| 1^{\n}_l , \tilde{0} \rangle
\langle \tilde{p}, p|  \tilde{q} ,q \rangle
\nonumber\\
\label{Ts1}
\eea
e
\be
\left| 1^{\m}_{l} \right\rangle ~= ~A^{\m \dagger}_{l} \left| 0 \right\rangle.
\label{field1}
\ee
Aqui, consideramos a normaliza\c{c}\~{a}o usual dos estados de momento em um
volume $V_{24}$ no espa\c{c}o transverso
\bea
\left.\left\langle p \right| q \right\rangle &~=~& 2 \pi 
\delta^{(24)} (p-q) \label{normstate1}\\
(2\pi )^{24}\delta^{(24)}(0) &~=~& V_{24}.
\label{normstate2}
\eea
Resulta, ap\'{o}s uma simples \'{a}lgebra, a contribui\c{c}\~{a}o dos
osciladores para o elemento de matriz
\bea
\left\langle\!\left\langle X^{\m}(\b_T)\left| K^{\rho} \right| X^{\m}(\b_T )
\right\rangle\!\right\rangle
~&=&~{\mbox{termos do c.m.}} 
-2 \a ' (2 \pi)^{(48)} \delta^{\m \n} \delta^{(24)}(p-q)\delta^{(24)}
(\tilde{p}-\tilde{q}) \times 
\nonumber\\
& &\sum_{n>0}\f{1}{n}\cos^2 n\s [\log (\tanh \th_n )^2\delta^{\rho \n } -
\delta^{\rho \rho }\sum_{k>0} \delta_{k k}].
\label{matrelemCM1}
\eea
Por sua vez os termos $CM$ pode ser divido em duas partes: uma contendo
somente operadores de posi\c{c}\~{a}o e \ momento e outra com a
contribui\c{c}\~{a}o dos osciladores. Usan-do a rela\c{c}\~{a}o de completeza
dos autoestados dos operadores de momento junto com os elementos de matriz
\be
\left. \left\langle x \right| p \right\rangle ~=~(2\pi \hbar ) ^{-12}
e^{i p \cdot x / \hbar},
\label{matrixxp1}
\ee
os termos contendo posi\c{c}\~{a}o e momento do $CM$ podem ser calculados.
A contribui\c{c}\~{a}o devida aos osciladores \'{e} obtida expressando os
osciladores em $T=0$ em termos dos osciladores em $T\neq0$ ou
escrevendo o v\'{a}cuo t\'{e}rmico em termos de v\'{a}cuo a temperatura 
nula. Os dois modos de c\'{a}lculo levam ao mesmo resultado. Usan\-do as
propriedades dos operadores de Bogoliubov mostra-se que as
contribui\c{c}\~{o}es dos termos onde h\'{a} mistura de operadores de centro
de massa com parte dos osciladores s\~{a}o cancelados. Os termos diferentes de
zero s\~{a}o todos proporcionais a $<0(\beta_{T})\left\vert K^{P}\right\vert
0(\beta_{T})>$. A rela\c{c}\~{a}o final para a entropia, levando em conta
todas  as contribui\c{c}\~{o}es \'{e}
\bea
& &\left\langle \!\left\langle X^{\mu}(\b_T)\left| K^{\rho} \right| 
X^{\m}(\b_T )\right\rangle\!\right\rangle
~=~ \nonumber\\
&- &(2\pi \hbar)^{-24}\left[ (2\pi\hbar)^{24}(2\a ' \t)^2 p^{\m} p'^{\n}
\delta^{(24)}(p-p') 
+ 
2\a ' \t (I^{\m}_2p'^{\n} + I'^{\n}_2 p^{\m}) + 
I^{\m}_2I^{\n}_2\prod_{j \neq \m , \n}I^j_1 \right]\nonumber\\ 
&\times &\delta^{(24)}(\tilde{p} - \tilde{p'})\sum_{m=1} 
\left[{\mbox n}^{\rho}_m 
\log {\mbox  n}^{\rho}_m + (1- {\mbox  n}^{\rho}_m) 
\log( 1- {\mbox  n}^{\rho}_m ) 
\right]
- 2 \a ' (2 \pi)^{(48)} \delta^{\m \n} \delta^{(24)}(p-p')\nonumber\\
&\times &\delta^{(24)}(\tilde{p}-\tilde{p'})  
\sum_{n>0}\f{1}{n}\cos^2 n\s \left[ \log (\tanh \th_n )^2\delta^{\rho \n } -
\delta^{\rho \rho }\sum_{k>0} \delta_{k k} \right],
\label{entropyRho1}
\eea
onde as integrais unidimensionais no dom\'{\i}nio $x \in [x_{0,}x_{1}]$ s\~{a}o
\bea
I_1 &~=~& -i \hbar (p'-p)^{-1}\left[ e^{\f{i}{\hbar}(p'-p)x_1} - 
e^{\f{i}{\hbar}(p'-p)x_0}\right]
\label{int11}\\
I_2 &~=~& -i \hbar (p'-p)^{-1}\left[  -i\hbar I_1 + 
x_1e^{\f{i}{\hbar}(p'-p)x_1} 
- x_0 e^{\f{i}{\hbar}(p'-p)x_0}\right].
\label{int21}
\eea 
e os estados de momento final e inicial s\~{a}o escolhidos por 
$| p>$ e $| p'>$, respectivamente. O n\'{u}mero de
excita\c{c}\~{o}es da corda no v\'{a}cuo t\'{e}rmico \'{e}
\be
{\mbox n}^{\rho}_{m} ~=~ \left\langle \!\left\langle 0(\b_T) \left| 
A^{\rho \dagger}_m A^{\rho}_m \right| 0(\b_T) \right\rangle\!\right\rangle 
= \sinh^2 \th_m \label{Nnumber1} 
\ee 

Uma vez que as condi\c{c}\~{o}es de contorno s\~{a}o imposta nas
coordenadas da folha mundo, podemos de forma semelhante obter express\~{o}es
para as entropias das cordas submetidas as demais $c.c$ $DD,DN$ e $ND$. Nestes
casos n\~{a}o existem operadores associados com as coordenadas e momentos de
centro de massa, mas vetores de posi\c{c}\~{a}o constante associados as suas
extremidades, n\~{a}o havendo contribui\c{c}\~{a}o destes termos para a entropia.

Os termos dos elementos de matriz diferentes de zero obtidos s\~{a}o
\bea 
{\mbox DD}&:& 
\left\langle \!\left\langle X^{\mu}(\b_T)\left| K^{\rho} \right| 
X^{\m}(\b_T )\right\rangle\!\right\rangle
=
2 \a ' (2 \pi)^{(48)} \delta^{\m \n} \delta^{(24)}(p-p')\delta^{(24)}
(\tilde{p}-\tilde{p'})  
\nonumber\\
&\times &
\sum_{n>0}\f{1}{n}\sin^2 n\s \left[ \log (\tanh \th_n)^2\delta^{\rho \n } -
\delta^{\rho \rho }\sum_{k>0} \delta_{k k} \right]
\label{entrDD1}\\
{\mbox DN}&:&
\left\langle \!\left\langle X^{\mu}(\b_T)\left| K^{\rho} \right| 
X^{\m}(\b_T )\right\rangle\!\right\rangle
= 
2 \a ' (2 \pi)^{(48)} \delta^{\m \n} \delta^{(24)}(p-p')\delta^{(24)}
(\tilde{p}-\tilde{p'})\!  
\nonumber\\
&\times &
\sum_{r=\ZZ + 1/2 }\!\f{1}{r}\sin^2 r\s \left[ \log (\tanh \th_r)^2
\delta^{\rho \n } -
\delta^{\rho \rho }\sum_{k>0} \delta_{k k} \right]
\label{entrDN1}\\
{\mbox ND}&:&
\left\langle \!\left\langle X^{\mu}(\b_T)\left| K^{\rho} \right| 
X^{\m}(\b_T )\right\rangle\!\right\rangle
= 
2 \a ' (2 \pi)^{(48)} \delta^{\m \n} \delta^{(24)}(p-p')
\delta^{(24)}(\tilde{p}-\tilde{p'}) 
\nonumber\\
&\times & \!\sum_{r=\ZZ + 1/2}\!\f{1}{r}\cos^2 r\s \left[ \log (\tanh \th_r)^2
\delta^{\rho \n } -
\delta^{\rho \rho }\sum_{k>0} \delta_{k k} \right]
\label{entrND1}
\eea
$(Z+\frac{1}{2})$, s\~{a}o n\'{u}meros inteiros. As express\~{o}es obtidas
d\~{a}o a entropia como fun\c{c}\~{a}o da folha mundo. Esta entropia n\~{a}o
pode ser pensada como a entropia do v\'{a}cuo da corda bos\^{o}nica que \'{e}
\ dada como a soma em todas as condi\c{c}\~{o}es espa\c{c}o - temporais da
entropia dos bosons escalares sem massa e n\~{a}o dependentes das c.c.

A contribui\c{c}\~{a}o para entropia dos estados da corda com
condi\c{c}\~{o}es de contorno DD, DN e ND pode ser calculada 
truncado as tr\^es rela\cc\~oes anteriores 
ap\'{o}s o primeiro termo de oscilador \ ou calculado o elemento de
matriz $K^{\rho}$ nos estados t\'{e}rmicos que descrevem campos de massa nula.
Os campos de massa nula formam um multipleto $U(1)$, $A^{j}%
=\alpha_{-1}^{j}=\left\vert 0\right\rangle $ onde $j=1,\ldots,p$ e um conjunto
de $(24-p)$ \ escalares $\phi^{a}=\alpha_{-1}^{a}\left\vert 0\right\rangle $
onde $a=p+1,\ldots,24$, dessa forma para os estados
\be 
\left.\left| \Psi^{\l} (\b_T) \right\rangle \!\right\rangle 
~=~\a^{\l}_{-1}(\b_T )\left. \left| 0 (\b_T ) \right\rangle \!\right\rangle
\label{fieldsatT1}
\ee
multiplicados pelas fun\c{c}\~{o}es dependentes de $\tau$ e $\sigma$
convenientes, respeitadas as condi\c{c}\~{o}es de contorno, calculamos os
elementos de matriz de $K^{\rho}$. Obtemos como express\~{a}o da entropia
$E=E_{\{A\}}+E_{\{ \phi \}}$
\bea 
E_{ \{ A \} } &~=~& 2\a ' p ( - \sin^2 \s \sum_{n=1}\log \cosh^2 \th_n 
+ 4 \cos^2 \s \sum_{r \in \ZZ+ 1/2} \log \cosh^2 \th_r ),
\nonumber\\
\label{entru11}\\
E_{ \{ \phi \} } &~=~& 2\a ' (24-p) ( - \sin^2 \s \sum_{n=1}\log \cosh^2 
\th_n + 4 \cos^2 \s \sum_{r \in \ZZ+ 1/2} \log \cosh^2 \th_r )
\nonumber\\
\label{entrscal1}.
\eea
Nas duas rela\c{c}\~{o}es acima o primeiro termo representa a
contribui\c{c}\~{a}o do setor DD e o segundo a contribui\c{c}\~{a}o dos
setores DN e ND. Somente o termo \ de entropia com as c.c NN depende de
$\hbar$. No limite semi-cl\'{a}ssico $\hbar \rightarrow 0$ a contribui\cc\~ao devida unicamente aos momenta \'e irrelevante. O termo dominante \'e o mesmo que domina no limite de tens\~ao infinita quando $\a' \rightarrow 0$. A entropia da corda devida as c.c. DD, DN e ND anula-se.

\section{Estados da Corda Bos\^onica T\'ermica no Espa\cc o AdS}

Para obter em primeira ordem a corda bos\^{o}nica t\'{e}rmica, vamos
aplicar a DCT \`a corda quantizada semicl\'{a}ssica descrita na
cap\'itulo 1. Vamos discutir o ansatz da DCT e o v\'{a}cuo t\'{e}mico
$\left\vert 0(\beta_{T})\right\rangle \ra $. No c\'{a}lculo da fun\c{c}\~{a}o de
parti\c{c}\~{a}o $Z(\beta_{T})$ h\'{a} diferen\c{c}as formais entre trabalhar
na espa\c{c}o de Hilbert total $\hat{\HH}$ e nos subspa\c{c}os f\'{\i}sicos $\HH$ e
$\tilde{\HH}$. Da forma de $Z(\beta_{T})$ em $\HH$, concluimos que operador de
Bogoliubov \'{e} conhecido e a termaliza\c{c}\~{a}o \'{e} vi\'{a}vel. Por
termaliza\c{c}\~{a}o, entendemos o processo de colocar o sistema em contato
com seu reservat\'{o}rio t\'{e}rmico de calor, o sistema inicial a temperatura
zero \'{e} levado a $T\neq0$. A intera\c{c}\~{a}o espec\'{\i}fica \'{e}
descrita via operador de Bogoliubov que mistura o par de osciladores. O
resultado desse procedimento \'{e} o aparecimento de dois novos graus de
liberdade t\'{e}rmicos. Diremos que o sistema est\'{a} dobrado quando expresso
em termos dos osciladores f\'{\i}sicos e os do reservat\'{o}rio
correspondentes, considerando uma temperatura determinada. O ansatz
fundamental da DCT \'{e} expresso pelo valor m\'{e}dio de um operador $O$ qualquer.
\be
\langle O \rangle = Z^{-1}(\b_T)\mbox{Tr}\left[ e^{-\b_T H} O \right] \equiv
\langle\langle 0(\b_T)|O|0(\b_T)\rangle\rangle,
\label{DCTansatz1}
\ee
Na aplica\c{c}\~{a}o da DCT o ansatz ser\'{a} modificado, quando
considerarmos uma teoria de cordas. O novo ansatz adotado \'{e} 
\be
\langle O \rangle = Z^{-1}(\b_T )\mbox{Tr}\left[ \delta(P=0) e^{-\b_T H}O\right]\equiv
\langle\langle 0(\b_T)|O| 0(\b_T )\rangle\rangle ,
\label{modifiedansatz11}
\ee
que
tamb\'{e}m, respeita a invari\^{a}ncia por reparametriza\c{c}\~{o}es na folha
mundo. Primeiro as simetrias da corda s\~{a}o fixada e ap\'{o}s o novo ansatz
\'{e} imposto, somente os estados f\'{\i}sicos contribuiem no c\'{a}lculo do
tra\c{c}o. Todo o conjunto de v\'{\i}nculos deve ser implementado antes da
imposi\c{c}\~{a}o desse novo ansatz. O v\'{a}cuo t\'{e}rmico na DCT tem a forma
\be
|0(\b_T )\rangle\rangle = \sum\limits_{w}\sum\limits_{\overline{w}}f_{w,\overline{w}}(\b_T)|w\rangle|\overline{w}\rangle ,
\label{thermvacexp1}
\ee  
onde neste caso $w$ e $\overline{w}$ s\~{a}o multi-\'{\i}ndices
correspondentes aos modos $\alpha$ e aos modos auxiliares $\beta$ do
reservat\'{o}rio, respectivamente, introduzimos a nota\c{c}\~{a}o para os
auto-valores dso operadores n\'{u}mero
\be
N_n = n \sum\limits_{\m = 1}^{D-1}k^{\m}_{n}~~,~~
\overline{N}_n = n \sum\limits_{\m = 1}^{D-1}\overline{k}^{\m}_{n},
\label{Neigenvalues11}
\ee
onde onde $k_{n}^{\mu}$ e $\bar{k}_{n}^{\mu}$ s\~{a}o auto-valores dos operadores
n\'{u}mero
\be
N^{\m}_{n}| \cdots k^{\m}_{n} \cdots \rangle = n k^{\m}_{n} | \cdots k^{\m}_{n} \cdots \rangle ,
~~,~~
\overline{N}^{\m}_{n}| \cdots \overline{k}^{\m}_{n} \cdots \rangle = n \overline{k}^{\m}_{n} | \cdots \overline{k}^{\m}_{n} \cdots \rangle ,
\label{Neigenvalues21}
\ee
respectivamente, para qualquer $\mu=1,2,\ldots,D-1$ e $n=1,2,\ldots,$ ou seja eles satisfazem as rela\c{c}\~{o}es
\bea 
f^{*}_{w',\overline{w}'}(\b_T)f_{w,\overline{w}}(\b_T) & = &
Z^{-1}(\b_T)\d(w',w)\d(\overline{w}',\overline{w})
\frac{\exp\left(\b_T \pi m^2 {\a '}^2 \right)}
{\left[ 1 - \exp \left( -\frac{\b_T \pi m {\a '}^2}{l} \right) \right]^{D-1}} \times
\nonumber\\
& & \int\limits_{-1/2}^{+1/2} ds \exp\left[ 2\pi \a ' \sum\limits_{n}
\left( \overline{\l}_n(\b_T ,s)\overline{N}_n + {\l}_n(\b_T ,s) N_n \right)\right],
\nonumber\\
\label{orthogf1}
\eea
onde $\delta(w,w')$ e $\delta(\overline{w},\overline{w}')$ s\~{a}o
nota\c{c}\~{o}es abreviadas para o produto de fun\c{c}\~{o}es delta, para cada
par de \'{\i}ndices no multi - \'{\i}ndice correspondente e
\be
{\l}_n(\b_T ,s) = - \b_T \o_n - \f{is}{\a '}~~,~~
\overline{\l}_n(\b_T ,s) = -\b_T \o_n + \f{is}{\a '} ~.
\label{lambdas1}
\ee
O v\'{\i}nculo pode ser escrito usando a representa\c{c}\~{a}o anal\'{\i}tica
da fun\c{c}\~{a}o delta
\be
\delta(P=0)\equiv\delta(\overline{N} - N) = \int\limits^{+1/2}_{-1/2} ds \,
e^{2\pi i s\left(\overline{N} - N\right)}.
\label{deltaanalytic1}
\ee
A rela\c{c}\~{a}o de ortogonalidade (\ref{orthogf1}) mostra que, como no caso do
espa\c{c}o - tempo de Minkowski, os coeficientes na expans\~{a}o do v\'{a}cuo
t\'{e}rmico s\~{a}o vetores do espa\c{c}o de Hilbert id\^{e}nticos aos do
espa\c{c}o de Hilbert das cordas, ou seja, o espa\c{c}o de Hilbert \^{H} que
tem os graus de liberdade do reservat\'{o}rio, tamb\'{e}m, como mostra a
rela\c{c}\~{a}o (*) na expans\~{a}o de $\left\vert 0(\beta_{T})\text{
}\right\rangle $ no espa\c{c}o de Hilbert total $\hat{\HH}=\HH\bigotimes
\widetilde{\HH,}$ estes vetores s\~{a}o escritos com os funcionais de Columbeau
\cite{c}, $i$ \'{e} a raiz quadrada de fun\c{c}\~{a}o delta. Isto sugere que o
v\'{a}cuo t\'{e}rmico \'{e} realmente um estado do espa\c{c}o f\'{\i}sico
total $\widehat{\HH}_{f\acute{\imath}sico}$ e $\widehat{\HH}$ e n\~{a}o de todo
espa\c{c}o. Al\'{e}m disso, n\~{a}o fator com fun\c{c}\~{a}o delta se o
tra\c{c}o de (\ref{DCTansatz1}) \ \'{e} tomada sobre $\widehat{\HH}_{f\acute{\imath}sico}$ em
vez de $\hat{\HH}$ e, consequentemente, n\~{a}o \ h\'{a} depend\^{e}ncia do
v\'{a}cuo t\'{e}rmico com os v\'{\i}nculos. por simplicidade, vamos trabalhar
no que depende do espa\c{c}o f\'{\i}sico. Ent\~{a}o a rela\c{c}\~{a}o (\ref{thermvacexp1})
pode ser expressa
\bea
|0(\b_T )\rangle\rangle & = &
Z^{-\frac{1}{2}}(\b_T)
\d(w',w)\d(\overline{w}',\overline{w})
        \frac{\exp 
                  \left( 
                        \frac{\b_T \pi m^2 {\a '}^2}
                             {2} 
                   \right)
             }
             {\left[ 1 - \exp \left( 
                                    -\frac{\b_T \pi m {\a '}^2}
                                          {l} 
                               \right)
              \right]^{\frac{D-1}{2}}
             }\times \nonumber\\
& & \sum\limits_{w}\sum\limits_{\overline{w}}
\exp\left[ -\b_T \pi \a ' \sum\limits_{n=1}^{\infty} \o_n \left( \overline{N}_n + N_n \right)\right]
|w,\overline{w}\rangle
\widetilde{|w,\overline{w}\rangle}.
\label{thermvacphs1}
\eea
Aqui, os estados de multi-\'{\i}ndice s\~{a}o $\left\vert
w,\widetilde{w}\ \ \right\rangle \in \widetilde{\HH}_{f\acute{\imath}sico}%
,$respectivamente. A fun\c{c}\~{a}o de parti\c{c}\~{a}o pode ser obtida
impondo que o v\'{a}cuo t\'{e}rmico \'{e} normalizado e tomando o tra\c{c}o do
operador identidade no subespa\c{c}o f\'{\i}sico
\be
Z(\b_T) = 
\frac{\exp \left( \b_T \pi m^2 {\a '}^2 \right)}
{\left[ 1 - \exp \left(-
                       \frac{ \b_T \pi m {\a '}^2}
                            {l} 
                 \right)
              \right]^{D-1}
             }
\prod\limits_{n=1}^{\infty}
\left( 1- e^{-\b_T \pi \a ' n \o_n} \right)^{2(1-D)}.
\label{partfuncphys1}
\ee             
Aqui, o fator 2 na exponencial vem das contribui\c{c}\~{o}es iguais dos
osciladores com $\alpha$ e $\beta$, respectivamente.

As rela\c{c}\~{o}es anteriores mostram que a decomposi\c{c}\~{a}o do
v\'{a}cuo t\'{e}rmico em termos dos estados f\'{\i}sicos \'{e} semelhante a do
campo livre qu\^antico no espa\c{c}o - tempo de Minkowski. Todavia, h\'{a} duas
diferen\c{c}as \ importantes. A primeira nas contribui\c{c}\~{o}es no modo
zero e o quadrado da massa que aparece na exponencial. A segunda
diferen\c{c}a, diz respeito a validade desse estado como estado de v\'{a}cuo
t\'{e}rmico de corda - \'{e} v\'{a}lido somente localmente no sistema de
refer\^{e}ncia de centro de massa, e \'{e}, ao longo de geod\'{e}sicas no
espa\c{c}o-tempo AdS.

O mapeamento da teoria em $T=0$ para $T\neq0$ \'{e} gerado pelo
operador de Bogoliubov dependente da temperatura, correspondendo a todos os
osciladores da sistema total
\be
\GG = G_0 + G + \overline{G},
\label{Bogoliubovopcf1}
\ee
onde o operador de Bogoliubov para o modo zero \'{e}
\be
G_0 = - i \theta_0 (\b_T )\sum\limits_{\m = 1 }^{D-1}
\left( \tilde{\a}^{\m}_{0} \a^{\m}_{0} - \a^{\dagger\m}_{0}\tilde{\a}^{\dagger\m}_{0} \right),
\label{Bogoliubovzero1}
\ee
e o par\^{a}metro $\theta_{0}$ \'{e} relacionado a fun\c{c}\~{a}o distribui\c{c}\~{a}o como
\be
\cosh \theta_{0}(\b_T ) = \left( 1 - e^{-\b_T \o_0} \right)^{-\f{1}{2}}.
\label{zeromodedistrib1}
\ee
A frequ\^encia do modo zero \'{e}
\be
\o_0 = \frac{\pi m {\a '}^2}{l}.
\label{zerofreq1}
\ee
Os operadores de Bogoliubov $G$ e $\overline{G}$ para os modos $\alpha$ e
$\beta$, respectivamente, tem a forma
\bea
G_{0} & = & \sum\limits_{n=1}^{\infty} G_n = -i\sum\limits_{n=1}^{\infty} \theta_n (\b_T )
\sum\limits_{\m = 1 }^{D-1}
\left( \tilde{\a}^{\m}_{n} \a^{\m}_{n} - \a^{\dagger\m}_{n}\tilde{\a}^{\dagger\m}_{n} \right),
\label{Gop1}\\
\overline{G}_{0} & = & \sum\limits_{n=1}^{\infty} \overline{G}_n = -i\sum\limits_{n=1}^{\infty} \overline{\theta}_n (\b_T )
\sum\limits_{\m = 1 }^{D-1}
\left( \tilde{\b}^{\m}_{n} \b^{\m}_{n} - \b^{\dagger\m}_{n}\tilde{\b}^{\dagger\m}_{n} \right). 
\label{barGop1}
\eea
Os coeficientes $\theta_{n}(\beta_{T})=\overline{\theta_{n}}(\beta_{T})$
s\~{a}o iguais para todos $n=1,2,\ldots$ $\mu=1,2,\ldots,D-1$, uma vez que os
osciladores s\~{a}o id\^{e}nticos em ambos os setores e ao longo de todas
dire\c{c}\~oes transversais do espa\c{c}o tangente. Suas rela\c{c}\~{o}es com as
distribui\c{c}\~{o}es bos\^{o}nicas s\~{a}o
\be
\cosh \theta_{n}(\b_T ) = \cosh \overline{\theta}_{n}(\b_T ) = \left( 1 - e^{-\b_T \o_n} \right)^{-\f{1}{2}},
\label{nmodedistrib1}
\ee
onde
\be
\o_n = \overline{\o}_n = \pi \a ' n \left( \frac{\O^{2}_{n} + 1}{\O_n} \right).
\label{nfreq1}
\ee
O v\'{a}cuo t\'{e}rmico da corda bos\^onica \'{e} a imagem do v\'{a}cuo total \`a temperatura zero 
\be
|0\rangle\rangle \equiv |0\rangle \tilde{|0\rangle} =
|0\rangle\rangle_0 \otimes_{n} |0\rangle\rangle_n \otimes_{n}\overline{|0\rangle\rangle}_n ,
\label{totalvac1}
\ee
sob a transforma\c{c}\~{a}o unit\'{a}ria gerada pelo operador de Bogoliubov
\be
|0(\b_T ) \rangle\rangle = e^{-i\GG}|0\rangle\rangle.
\label{thermvacmap1}
\ee 
Como $\left\vert 0\right\rangle \rangle$ pertence ao espa\c{c}o f\'{\i}sico
total, o operadore de Bogoliubov mapea $\widehat{\HH}_{f\acute{\imath}sico}$ no
espa\c{c}o de Hilbert t\'{e}rmico $\ \widehat{\HH}$ $(\beta_{T})$. O v\'{a}cuo
total \'{e} aniquilado por todos os operadores de aniquila\c{c}a\~{o} e tem
invari\^{a}ncia translacional. O v\'{a}cuo t\'{e}rmico pode ser definido da mesma forma se os
operadores s\~{a}o construidos com a a\c{c}\~{a}o sobre o conjunto de todos operadores
\be
\OO \equiv \{ O \} =
\{ 
\a^{\m}_{0}, \a^{\dagger\m}_{0}, \tilde{\a}^{\m}_{0}, \tilde{\a}^{\dagger}_{0};
\a^{\m}_{n}, \a^{\dagger\m}_{n}, \tilde{\a}^{\m}_{n}, \tilde{\a}^{\dagger}_{n};
\b^{\m}_{n}, \b^{\dagger\m}_{n}, \tilde{\b}^{\m}_{n}, \tilde{\b}^{\dagger}_{n}
\},
\label{setosczeroT1}
\ee
de transforma\c{c}\~{o}es similares geradas pelo operador de Bogoliubov
\be
\OO (\b_T ) = e^{-i\GG}\OO e^{i\GG} =  \{ e^{-i\GG} O e^{i\GG} \}.  
\label{setoscnonzeroT1}
\ee
O espa\c{c}o $\HH_{(\beta_{T})}$ tem a estrutura de um espa\c{c}o de Fock. Os
estados de v\'{a}cuo t\'{e}rmico satisfazem as rela\c{c}\~{o}es
\bea
\a^{\m}_{0} (\b_T ) |0(\b_T ) \rangle\rangle & = & 
\a^{\m}_{n} (\b_T ) |0(\b_T ) \rangle\rangle =
\b^{\m}_{n} (\b_T ) |0(\b_T ) \rangle\rangle = 0,
\label{firstthermvacrel1}\\
\tilde{\a}^{\m}_{0} (\b_T ) |0(\b_T ) \rangle\rangle & = & 
\tilde{\a}^{\m}_{n} (\b_T ) |0(\b_T ) \rangle\rangle =
\tilde{\b}^{\m}_{n} (\b_T ) |0(\b_T ) \rangle\rangle = 0.
\label{secondthermvacrel1}
\eea
Como o operador de Bogoliubov mistura os modos de osciladores a
temperatura zero de todos os setores, a temperatura finita osciladores sem til
e com til n\~{a}o representam mais grau de liberdade da corda e do
reservat\'{o}rio, respectivamente. Portanto, eles representam as
oscila\c{c}\~{o}es t\'{e}rmicas do sistema aquecido que resulta da
intera\c{c}\~{a}o a temperatura zero da corda e seu reservat\'{o}rio. Um
estado de corda t\'{e}rmica, dever\'{a} conter um n\'{u}mero arbitr\'{a}rio de
excita\c{c}\~{o}es de todos os setores e tem como forma geral
\bea
& & |\Psi^{\m_1 \ldots \n_1 \ldots \rho_1 \ldots \t_1}_{m_1 \ldots \n_1 \ldots p_1 \ldots q_1}(\b_T)\rangle\rangle =\nonumber\\
& & \left[ \a^{\dagger\m_1}_{m_1}(\b_T) \right]^{k^{\m_1}_{m_1}} \cdots
\left[ \b^{\dagger\n_1}_{n_1}(b_T) \right]^{\overline{k}^{\n_1}_{n_1}} \cdots
\left[ \tilde{\a}^{\dagger\rho_1}_{p_1}(\b_T) \right]^{s^{\rho_1}_{p_1}}
\cdots
\left[ \tilde{\b}^{\dagger\t_1}_{q_1}(\b_T) \right]^{\overline{s}^{\t_1}_{q_1}}
\cdots
|0(\b_T )\rangle\rangle.
\nonumber\\
\label{genthermstate1}
\eea
O estado cont\'{e}m $k_{m_{1}}^{\mu_1}$ exita\c{c}\~{o}es t\'{e}rmicas do tipo
$\alpha_{m_{1}}$ na dire\c{c}\~{a}o $\mu_{1}$, $\bar{k}_{n_{1}}^{\n_{1}}$
excita\c{c}\~{o}es t\'{e}rmicas do tipo $\beta_{n_{1}}$ na dire\c{c}\~{a}o
$\gamma_{1}$, etc.

As simetrias da corda t\'{e}rmica podem ser verificadas utilizando-se a
\'{a}lgebra conforme nos estados t\'{e}rmicos. Todavia, os operadores $L_{n}$
e $\bar{L_{n}}$ n\~{a}o comutam com operadore de Bogoliubov a \'{a}lgebra
conforme \'{e} quebrada a temperatura finita. \'{E} natural perguntamos se
h\'{a} simetrias e v\'{\i}nculos que possam ser impostos sobre os estados da
corda. A resposta a quest\~{a}o \'{e} obtida notando que a din\^{a}mica da
corda a temperatura finitas pose ser derivada da Lagrangeana
\be 
\LL_{2}(\b_T ) = e^{-i\GG}\hat{\LL}_{2}e^{i\GG},
\label{fintemplagr1}
\ee
onde $\hat{\LL}_2 = \LL_2 - \tilde{\LL}_2$ e $\LL_2$ \'{e} a Lagrangeana
correspondente a a\c{c}\~{a}o truncada dada pela rela\c{c}\~{a}o (\ref{fintemplagr1}) e
$\tilde{L}_{2}$ \'{e} a parte do reservat\'{o}rio associado. Da
Lagrangeana acima, a Hamiltomiana e momento na folha mundo em $D = 2+1$ s\~{a}o
\be
\hat{H} = H - \tilde{H}~~,~~\hat{P} = P - \tilde{P},
\label{tothamtotmom1}
\ee
e
mostra que as rela\c{c}\~{o}es de comuta\c{c}\~{a}o s\~{a}o
\be
\left[ \hat{H}, \GG \right] = \left[ \hat{P}, \GG \right] = 0.
\label{commhammom1}
\ee
Sendo $\widehat{H}$ a hamiltonia total da corda bos\^{o}nica podemos
interpretar \ $\widehat{P}$ como sendo o momento total. Finalmente, qual
estado f\'{\i}sico $\left\vert \Psi_{f\acute{\imath}sico}\right\rangle
\rangle=\left\vert \Psi_{f\acute{\imath}sico}\right\rangle $\ $\widetilde
{\left\vert \Psi_{f\acute{\imath}sico}\right\rangle }$\ pode ser mapeado no
estado t\'{e}rmico
\be
|\Psi_{fisico}(\b_T ) \rangle\rangle = e^{-i\GG} | \Psi_{fisico}\rangle\rangle,   
\label{fisicostatesatT1}
\ee
que \'{e} invariante por transla\c{c}\~{a}o na folha mundo
\be
\hat{P}|\Psi_{fisico}(\b_T ) \rangle\rangle = 0.
\label{invarfisico1}
\ee
A rela\c{c}\~{a}o acima juntamente com a invari\^{a}ncia da Hamiltoniana
s\~{a}o usadas para definir os estados t\'{e}rmicos. Observamos que na
rela\c{c}\~{a}o acima, os operadores est\~{a}o a temperatura zero e o estado a
temperatura finita.

Considerando os modos de oscila\c{c}\~{a}o da corda no espa\c{c}o de
Hilbert f\'{\i}sico, a entropia pode ser localmente definida como o valor
esperado do operador $K$ no v\'{a}cuo t\'{e}rmico da corda, como mostrado no cap\'itulo anterior. 
O operador $K$ \'{e}
\bea
K &=& - \sum\limits_{n=1}^{\infty}\sum\limits_{\m=1}^{D-1}
\left[ 
	\left( 
		\a^{\mu \dagger}_{n} \a^{\m}_{n} + \b^{\m\dagger}_{n} \b^{\m}_{n}
	\right)
				\log\sinh^2\theta_{n}(\b_T) -
	\left( 
		\a^{\m}_{n} \a^{\mu \dagger}_{n} + \b^{\m}_{n} \b^{\m\dagger}_{n} 
	\right)
				\log\cosh^2\theta_{n}(\b_T)			
\right]\nonumber\\
&-&
\sum\limits_{\m=1}^{D-1}
\left[ 
		\a^{\mu \dagger}_{0} \a^{\m}_{0} \log\sinh^2\theta_{0}(\b_T) -
		\a^{\m}_{0} \a^{\mu \dagger}_{0} \log\cosh^2\theta_{0}(\b_T)			
\right].
\label{entropyop1}
\eea 
Usando o valor esperado do operador n\'{u}mero no v\'{a}cuo t\'{e}rmico
\bea
\langle\langle 0 (\b_T ) | \a^{\mu \dagger}_{n} \a^{\m}_{n} |0 (\b_T ) \rangle\rangle =
\langle\langle 0 (\b_T ) | \b^{\mu \dagger}_{n} \b^{\m}_{n} |0 (\b_T ) \rangle\rangle =
\sinh^2 \theta_{n}(\b_T ),
\label{numbopexp1}
\eea
para todos osciladores, podemos escrever a entropia
\bea
S &=& 2(D-1)k_B \sum\limits_{n=1}^{\infty}
\left[
			\b_T \pi \a ' n \o_n f(\pi \a ' n \o_n ) + \log(1+f(\pi \a ' n \o_n))
\right]
\nonumber\\
&+&
(D-1)k_B \left[
							\b_T \frac{\pi m {\a '}^2}{l}f(\frac{\pi m {\a '}^2}{l}) + \log \left(1+
							 f(\frac{\pi m {\a '}^2}{l}) \right)
				 \right],	
\label{entropy11}
\eea
onde
\be
f(\o_n) = \frac{1}{e^{\b_T \o_n}-1},
\label{thetadistrib1}
\ee
\'{e} a fun\c{c}\~{a}o distribui\c{c}\~{a}o por $n=1,2,\ldots$. Por
defini\c{c}\~{a}o a energia livre \'{e} dada por valor esperado do operador
$F$ no v\'{a}cuo t\'{e}rmico, onde
\be
F = -\frac{1}{k_B}K + H.
\label{freeenergy1}
\ee
Com o uso dos c\'{a}lculos acima e a forma expl\'{\i}cita da Hamiltoniana
local, obtemos a energia livre de Helmholtz
\bea
\langle \langle F \rangle  \rangle &=& (D-1)\sum\limits_{n=1}^{\infty}
				 \left[ 	
				 4\pi \a ' n\o_n f(\pi \a ' n \o_n) + \frac{2\pi m {\a '}^2}{l}
				 f(\frac{\pi m {\a '}^2}{l})
				 \right]
\nonumber\\
&+&
\frac{(D-1)}{\b_T} \left[
				 2 \sum\limits_{n=1}^{\infty}
				 	\log\left(1 + f(\pi \a ' n \o_n)\right) +
				 	\log\left(1 + f(\frac{\pi m {\a '}^2}{l}) \right)
				 	\right]
				 	- \pi m^2 {\a '}^2
				 ,	
				 \nonumber\\			 
\label{freeenergycalc1}
\eea
onde os dois \'{u}ltimos termos representam a contribui\c{c}\~{a}o dos modos
zero e a da massa da corda, respectivamente. Os dois \'{u}ltimos termos na
entropia diferem a fun\c{c}\~{a}o $S_{0}$, com tens\~{a}o da corda e a
constante cosmol\'{o}gica. A temperatura constante e para
\be
T^{2}_{s} >> \f{m \b_T }{4 \pi} \sqrt{-\L},
\label{inftylimittension1}
\ee 
$S_{0}$ depende da temperatura como
\be
S_0 \approx 1  + \log\left( 1+ \frac{1}{\b_T \o_0} \right).
\label{approx11}
\ee
Assim, esta contribui\c{c}\~{a}o d\'{e}ve ser relevante a altas temperaturas
\be
T >> \frac{m}{4 \pi k_B}\sqrt{-\L}. 
\label{hightemp1}
\ee
Para valores, onde a tens\~{a}o da corda $T_s$ \'{e}
\be
T^{2}_{s} << \f{m \b_T }{4 \pi} \sqrt{-\L},
\label{zerolimittension1}
\ee 
os \'{u}ltimos termos na entropia dependem da temperatura como
\be
S_0 \approx \log(2 - \b_T \o_0) + \b_T \o_0 -(\b_T \o_0 )^2.
\label{approx21}
\ee
Neste casa, h\'{a} temperatura c\'{u}tica $Tc$
\be
T_c = \frac{8\pi k_B}{m T^{2}_{s}\sqrt{-\L}}.
\label{criticaltemp1}
\ee
Para \ $T < T_c$ o modo zero da entropia n\~{a}o \'{e} bem definido.
Este resultado pode ser interpretado como n\~{a}o validade do procedimento de
quantiza\c{c}\~{a}o semicl\'{a}ssico no limite de tens\~{a}o nula da teoria
de cordas. Neste limite, a intera\c{c}\~{a}o entre os osciladores da corda vai
al\'{e}m da aproxima\c{c}\~{a}o em primeira ordem da expans\~{a}o pertubativa
em $\in$. Os efeitos da corda s\~{a}o reduzidos no limite de tens\~{a}o grande
o suficiente, para que a corda se comporte mais como uma part\'{\i}cula.
Considera\c{c}\~oes semelhantes podem ser tiradas da energia livre.

\section{V\'acuo T\'ermico no Espa\cc o de Hilbert Total}

Aplicando o postulado fundamental da DCT do cap\'itulo 3, modificado para teoria de cordas como visto anteriormente,  o valor esperado de um operador $O$ no espa\c{c}o de Hilbert total \'e
\be
Z^{-1}(\b_T )\mbox{Tr}\left[ \delta(P=0) e^{-\b_T H}O\right]\equiv
\langle\langle 0(\b_T)|O| 0(\b_T )\rangle\rangle ,
\label{modifiedansatz112}
\ee
onde o estado do v\'{a}cuo t\'{e}rmico no espa\c{c}o de Hilbert total \'e
\bea
|0(\b_T ) \rangle\rangle & = & 
\frac{ Z^{-\frac{1}{2}}(\b_T)
\exp (\frac{\b_T \pi m^2 {\a '}^2)}{2}}
			{
				\left[ 
							1 - \exp
											\left( 
													-\b_T \frac{\pi m {\a '}^2}{l}
											\right)
				\right]^{\frac{D-1}{2}}
			} 
		\sum\limits_{w}
		\sum\limits_{\overline{w}}
	\left[
	\int\limits_{-1/2}^{+1/2} ds
		\exp 
			\left(
				i\pi n s \sum\limits_{\m=1}^{D-1} 
					\left(
					k^{\m}_{n} - \overline{k}^{\m}_{n}
					\right)
			\right)		
	\right]^{-\frac{1}{2}}
\nonumber\\
&\times &		
\exp 
		\left(
		-\b_T \pi \a ' \sum\limits_{n=1}^{\infty}n\o_n
									 \sum\limits_{\m=1}^{D-1}
									 \left(
									 k^{\m}_{n} + \overline{k}^{\m}_{n}
									 \right)	
		\right)	
|w,\overline{w}\rangle	\widetilde{|w,\overline{w}\rangle}.	
\label{fullthermvacuum1}
\eea
A fun\c{c}\~{a}o de parti\c{c}\~{a}o t\'{e}rmica tem forma explicita
\be
Z(\b_T ) = \frac{\exp (\b_T \pi m^2 {\a '}^2)}
			{
				\left[ 
							1 - \exp
											\left( 
													-\b_T \frac{\pi m {\a '}^2}{l}
											\right)
				\right]^{D-1}
			}
\int\limits_{-1/2}^{+1/2} ds \prod\limits_{n=1}^{\infty}
				\left[
					\left(
					1-e^{\pi \a ' n \l_n(\b_T,s)}
					\right)
				\left(
					1-e^{\pi \a ' n \overline{\l}_n(\b_T,s)}
					\right)
				\right]^{1-D}.
\label{fullHpartfunct1}
\ee
No espa\c{c}o de Hilbert total a Hamiltoniana tem a forma
\be
H' =  2\pi\a ' \sum\limits_{n \geq 1} 
\left[
	\left( 
				\frac{\O^{2}_{n} + 1}{\O_n} 
	\right)  
	\left( N_n + \overline{N}_{n} \right)
	+
	2\pi i s 
	\left( 
				N_n - \overline{N}_n 
	\right)
\right]
+\frac{\pi m {\a '}^2}{l} \a^{\dagger}_{0} \cdot \a_{0} - 
\pi m^2{\a '}^2 ,
\label{fullHam1}
\ee
que difere da (\ref{hamiltonian}) por conter a condi\cc\~ao de n\'iveis iguais (level matching) dada na express\~ao
(\ref{momentum}). A Hamiltoniana (\ref{fullHam1}) est\'{a} de acordo com a interpreta\c{c}\~{a}o do par\^{a}metro $s$ como multiplicador de Lagrange \cite{ng3}. A presen\cc a dos v\'inculos na Hamiltoniana (\ref{fullHam1}) se deve a expans\~ao do v\'acuo t\'ermico no espa\cc o de Hilbert total enquanto a Hamiltoniana (\ref{hamiltonian}) de partida tem os estados escolhidos no espa\cc o de Hilbert f\'isico. Desenvolver o c\'alculo das grandezas f\'isicas da corda bos\^onica em qualquer uma das representa\cc\~oes deve ser equivalente sendo que trabalhar no espa\c{c}o de Hilbert total implica manipular os funcionais de Columbeau \cite{c}.

A teoria de cordas bos\^onicas \'e conforme no espa\cc o AdS $D=2+1$ \cite{bhtz} onde a termaliza\cc\~ao no m\'etodo DCT \'e exata. Entretanto, a \'algebra de Virasoro n\~ao se realiza na representa\cc\~ao de estados  t\'ermicos aparecendo uma quebra da simetria conforme. \'E interessante estudar a termaliza\cc\~ao da \'algebra de Virasoro usando a representa\cc\~ao de osciladores que possibilitam implementar de modo direto o formalismo DCT. O passo inicial para a constru\cc\~ao da \'algebra de Virasoro t\'ermica foi dado utilizando t\'ecnicas alg\'ebricas de Wigner-Heisenberg na constru\cc\~ao dos geradores da \'algebra de Virasoro \cite{eg4}.

\newpage

\chapter{Conclus\~oes}

Calculamos a entropia e a energia livre para todas as solu\c{c}\~{o}es da
corda t\'{e}rmica bos\^{o}nica aberta , onde s\~{a}o levadas em considera%
\c{c}\~{a}o todas as condi\c{c}\~{o}es de contorno poss\'{\i}veis \cite{eg1}.

Formulamos uma teoria a temperatura finita para excita\c{c}\~{o}es t\'{e}%
rmicas livres da corda bos\^{o}nica fechada no espa\c{c}o AdS na abordagem
de DCT. A m\'{e}trica no espa\c{c}o AdS \'{e} tratada exatamente quando a
corda e o reservat\'{o}rio t\'{e}rmico s\~{a}o semi-classicamente
quantizados em teoria de perturba\c{c}\~{a}o at\'{e} primeira ordem com
respeito ao par\^{a}metro adimensional $\epsilon =\alpha ^{\prime }H^{2}$
onde $H$ \'{e} a constante de Hubble. Com fundo de buraco negro no AdS
conforme $D=2+1$, a quantiza\c{c}\~{a}o \'{e} exata. O m\'{e}todo pode ser
extendido a espa\c{c}o-tempo AdS arbitr\'{a}rio. A aproxima\c{c}\~{a}o \'{e}
tomada no sistema de refer\^{e}ncia de centro de massa, sendo justificada
pelo fato de que em primeira ordem a din\^{a}mica da corda \'{e} determinada
somente pela intera\c{c}\~{a}o entre os modos livres de oscila\c{c}\~{o}es
da corda e a solu\c{c}\~{a}o de fundo exata. A corda t\'{e}rmica bos\^{o}%
nica fechada em primeira ordem \'{e} obtida por termaliza\c{c}\~{a}o do
sistema em $T=0$ efetuada pelos operadores de Bogoliubov da DCT.
Determinamos os estados da corda t\'{e}rmica bos\^{o}nica fechada e
calculamos a entropia local e a energia livre nos sistema de refer\^{e}ncia
de centro de massa. Discutimos tamb\'{e}m a rela\c{c}\~{a}o entre a
Hamiltoniana no espa\c{c}o de Hilbert total e o espa\c{c}o de Hilbert f\'{\i}%
sico. A DCT tem-se mostrado prof\'{\i}cua neste procedimento can\^{o}nico e
perturbativo em que submetemos as cordas.

O pr\'{o}ximo passo no desenvolvimento de nosso trabalho indica caminhos na
aplica\c{c}\~{a}o a cordas supersim\'etricas e tamb\'{e}m ao estudo das $D$%
-branas. Uma poss\'{\i}vel generaliza\c{c}\~{a}o do m\'{e}todo apresentado
nesta tese, com aplica\c{c}\~{o}es na teoria de cordas e supercordas, \'{e}
a termaliza\c{c}\~{a}o no formalismo DCT das representa\c{c}\~{o}es da
\'algebra de Virasoro usando t\'{e}cnicas alg\'{e}bricas de Wigner-Heisenberg
para sistemas bos\^{o}nicos \cite{eg4}.


\newpage

\begin{thebibliography}{999}
\addcontentsline{toc}{chapter}{Refer\^encias}
\bibitem{mvm} 
  M.~A.~Vazquez-Mozo,
  Phys.\ Lett.\ B {\bf 388}, 494 (1996);
  J.~L.~F.~Barbon, M.~A.~Vazquez-Mozo,
  Nucl.\ Phys.\ B {\bf 497}, 236 (1997).

\bibitem{od} A. Bytsenko, S. Odintsov and L. Granada, Mod. Phys. Lett.
A{\bf 11}, 2525(1996).

\bibitem{rsz} J. Ambjorn, Yu. Makeenko, G. W. Semenoff and R. Szabo,
Phys. Rev. D{\bf 60}, 106009(1999).

\bibitem{kog1} G. Dvali, I. I. Kogan and M. Shifman, Phys. Rev.
D{\bf 62}, 106001(2000).

\bibitem{kog2} S. Abel, K. Freese and I. I. Kogan, JHEP {\bf 0101}, 039(2001).

\bibitem{kog3}
  I.~I.~Kogan, A.~Kovner and M.~Schvellinger,
  JHEP {\bf 0107}, 019 (2001).

\bibitem{rab1} S. A. Abel, J. L. F. Barbon, I. I. Kogan, E. Rabinovici,
JHEP {\bf 9904}, 015(1999).

\bibitem{rab2} J. L. F. Barbon, E. Rabinovici, JHEP {\bf 0106}, 029(2001).

\bibitem{gub}
  S.~S.~Gubser, S.~Gukov, I.~R.~Klebanov, M.~Rangamani and E.~Witten,
  J.\ Math.\ Phys.\  {\bf 42}, 2749 (2001)

\bibitem{gw}M. B. Green and P. Wai, Nucl. Phys. B{\bf 431}, 131(1994). 

\bibitem{bg}M. B. Green, Nucl. Phys. B{\bf 381}, 201(1992). 

\bibitem{gg}M. B. Green and M. Gutperle, Nucl.Phys. B{\bf 476}, 484(1996). 

\bibitem{ml1} M. Li, Nucl. Phys. B {\bf 460}, 351(1996).

\bibitem{ck} C. Callan Jr. and I. Klebanov, Nucl. Phys. B{\bf 465}, 473(1996). 

\bibitem{dv1} P. Di Vecchia  and A. Liccardo, {\it D-Branes in String Theory I}, 
hep-th/9912161; {\it D-Branes in String Theory II}, hep-th/9912275. 

\bibitem{dv2} P. Di Vecchia, M. Frau, A. Lerda and A. Liccardo, Nucl. Phys.
B{\bf 565}, 397(2000).
\bibitem{yl1} Y. Leblanc, Phys. Rev. D{\bf 38}, 3087(1988).

\bibitem{yl2} Y. Leblanc, Phys. Rev. D{\bf 36}, 1780(1987); Phys. Rev. 
D{\bf 37}, 1547(1988); Phys. Rev. D{\bf 39}, 1139(1989); Phys. Rev. D{\bf 39}, 3731(1989).

\bibitem{yl3} Y. Leblanc, M. Knecht and J. C. Wallet, Phys. Lett. B{\bf 237}, 357(1990).
\bibitem{fn1} H. Fujisaki and K. Nakagawa, Prog. Theor. Phys. 
{\bf 82}, 236(1989); 
Prog. Theor. Phys. {\bf 82}, 1017(1989); Prog. Theor. Phys. {\bf 83}, 18(1990); 
Europhys. Lett. {\bf 20}, 677(1992); Europhys. Lett. {\bf 28}, 471(1994). 

\bibitem{hf} H. Fujisaki, Il Nuovo Cimento, {\bf 108A}, 1079(1995).

\bibitem{fn2} H. Fujisaki and K. Nakagawa, Europhys. Lett. {\bf 35}, 493(1996).
\bibitem{ng1}
  D.~L.~Nedel, M.~C.~B.~Abdalla and A.~L.~Gadelha,
  Phys.\ Lett.\ B {\bf 598}, 121 (2004).
\bibitem{ng2}
  M.~C.~B.~Abdalla, A.~L.~Gadelha and D.~L.~Nedel,
  JHEP {\bf 0510}, 063 (2005).
\bibitem{ng3}
  M.~C.~B.~Abdalla, A.~L.~Gadelha and D.~L.~Nedel,
  Phys.\ Lett.\ B {\bf 613}, 213 (2005).
\bibitem{ng5}
  M.~C.~B.~Abdalla, A.~L.~Gadelha and D.~L.~Nedel,
  PoS({\bf WC2004}), 032 (2004).
\bibitem{ivv1}
  I.~V.~Vancea,
  Phys.\ Lett.\ B {\bf 487}, 175 (2000).
\bibitem{ivv2}
  M.~C.~B.~Abdalla, A.~L.~Gadelha and I.~V.~Vancea,
  Phys.\ Rev.\ D {\bf 64}, 086005 (2001).
\bibitem{ivv3}
  M.~C.~B.~Abdalla, A.~L.~Gadelha and I.~V.~Vancea,
  Phys.\ Rev.\ D {\bf 66}, 065005 (2002).
\bibitem{ivv4}
  M.~C.~B.~Abdalla, A.~L.~Gadelha and I.~V.~Vancea,
  {\em $D$-branes at finite temperature in TFD},
  hep-th/0308114.
\bibitem{ivv5}
  M.~C.~B.~Abdalla, A.~L.~Gadelha and I.~V.~Vancea,
  Int.\ J.\ Mod.\ Phys.\ A {\bf 18}, 2109 (2003).
\bibitem{ivv6}
  M.~C.~B.~Abdalla, A.~L.~Gadelha and I.~V.~Vancea,
  Nucl.\ Phys.\ Proc.\ Suppl.\  {\bf 127}, 92 (2004).
\bibitem{ivv8}
  I.~V.~Vancea,
  Phys.\ Rev.\ D {\bf 74}, 086002 (2006).
\bibitem{ivv9}
  I.~V.~Vancea,
  PoS({\bf IC2006}), 36(2006).
\bibitem{ivv10}
  I.~V.~Vancea,
  Phys.\ Rev.\ D {\bf 74}, 086002 (2006).
\bibitem{vesan}
  H.~J.~de Vega and N.~Sanchez,
  Phys.\ Lett.\ B {\bf 197}, 320 (1987).
\bibitem{ns2}
  H.~J.~de Vega and N.~Sanchez,
  Nucl.\ Phys.\ B {\bf 299}, 818 (1988).
\bibitem{nsa}
  N.~G.~Sanchez,
  Phys.\ Lett.\ B {\bf 195}, 160 (1987).
\bibitem{ns3}
  A.~L.~Larsen and N.~Sanchez,
  Phys.\ Rev.\ D {\bf 50}, 7493 (1994).
\bibitem{ns4}
  H.~J.~de Vega, A.~L.~Larsen and N.~G.~Sanchez,
  Phys.\ Rev.\ D {\bf 58}, 026001 (1998).
\bibitem{bhtz}
  M.~Banados, C.~Teitelboim and J.~Zanelli,
  Phys.\ Rev.\ Lett.\  {\bf 69}, 1849 (1992).
\bibitem{bhtz1}
  M.~Banados, M.~Henneaux, C.~Teitelboim and J.~Zanelli,
  Phys.\ Rev.\  D {\bf 48}, 1506 (1993)
\bibitem{hw}G.T. Horowitz and D.L. Welch, Phys. Rev. Lett. 71 (1993) 328.
\bibitem{ggrt}
P. Goddard, J. Goldstone, C. Rebi e C. B. Thorn, Nucl.\ Phys.\ {\bf B56}, 109(1973).
\bibitem{mt}T.~ Matsubara, Prog.\ Theor.\  Phys. {\bf 14}, 351 (1955).
\bibitem{tu1}Y. Takahashi, H. Umezawa, Collective\ Phenom.\ {\bf 2}, 55 (1975).
\bibitem{tu2}Y. Takahashi, H. Umezawa, Int.\ J.\ Mod.\ Phys.\ {\bf B10}, 1755 (1996).   
\bibitem{eg1}
  M.~C.~B.~Abdalla, E.~L.~Graca and I.~V.~Vancea,
  Phys.\ Lett.\ B {\bf 536}, 114 (2002).
\bibitem{eg2}
  H.~Belich, E.~L.~Graca, M.~A.~Santos and I.~V.~Vancea,
  JHEP {\bf 0702}, 037 (2007)
\bibitem{eg3}
  E.~L.~Graca and I.~V.~Vancea,
  {\em Thermal string vacuum in black-hole AdS spacetime},
  hep-th/0505210, submetido para publica\cc\~ao.
\bibitem{eg4}
  E.~L.~da Graca, H.~L.~Carrion and R.~de Lima Rodrigues,
  Braz.\ J.\ Phys.\  {\bf 33}, 333 (2003).
\bibitem{eg5}
E.~L.~Graca, I.~V.~Vancea, {\em Estados T\'ermicos da Corda Aberta no Formalismo DCT},
apresentado no encontro {\em 2nd International Conference on Fundamental Interactions, 
Domingos Martins, Espirito Santo, Brazil, 6-12 Jun 2004}(poster). 
\bibitem{hat} B.~Hatfield,
  {\em Quantum Field Theory of Point Particles and Strings}, Ed. Addison-Wesley (1992).
\bibitem{green} M.~B.~Green, J.~H.~Schwarz and E.~Witten, {\em Superstring Theory}, Vol.1, 
  Ed. Cambridge University Press (1987). 
\bibitem{polchinski} J.~Polchinski, {\em String Theory}, Vol. 1,Ed. Cambridge University Press (1998).
\bibitem{greiner}
  W.~Greiner and J.~Reinhardt, {\em Field Quantization}, Ed. Springer-Verlag (1983).
\bibitem{ubook}H. Umezawa, {\em Advanced Field Theory: Micro, Macro and Thermal Physics},  Ed. AIP New-York, (1993). 
\bibitem{c}J. F. Columbeau, {\em Elementary Introduction to New Generalized Functions}
Ed. North Holland, (1985).
\bibitem{tesewagner} W.~Paniago de Souza, {\em Corda Bos\^onica \`a Temperatura Finita}, tese de mestrado IFT-UNESP (2002), hep-th/0208134.
\bibitem{c} J.~ F.~ Columbeau, {\em Elementary Introduction to New Generalized Functions}
North Holland, 1985.
\bibitem{alg} A.~L.~Gadelha, {\em Dp-Branas \`a Temperatura Finita}, tese de doutorado IFT-UNESP, (2002). 



\end{thebibliography}
\end{document}